\documentclass[sigconf, nonacm]{acmart}
\usepackage{soul}
\usepackage[utf8]{inputenc}
\usepackage{xcolor}
\usepackage{algorithm}
\usepackage{algpseudocode}
\usepackage{listings}
\theoremstyle{definition}

\usepackage{graphicx}
\usepackage{subcaption}
\usepackage{caption}
\usepackage{algpseudocode}
\usepackage{array}
\usepackage{cleveref}
\makeatletter
\AtBeginDocument{%
\let\cref@old@makefntext\@makefntext%
\long\def\@makefntext{%
  \cref@constructprefix{footnote}{\cref@result}%
  \protected@edef\cref@currentlabel{%
    [footnote][\arabic{footnote}][\cref@result]%
    \p@footnote\@thefnmark}%
  \cref@old@makefntext}%
 } 
\makeatother
\usepackage{enumitem}
\usepackage{ifthen}
\usepackage[bottom]{footmisc}
\usepackage{amsmath}

\newtheorem{example}{Example}
\newcommand{\mparagraph}[1]{\paragraph{\textbf{#1}}}

\newcommand{\narrow}[1]{\ensuremath{\text{\scalebox{0.7}[1.0]{\textsf{#1}}}}}
\newcommand{\Space}[1]{\ensuremath{\mathbf{#1}}}
\newcommand{\Expectation}{\ensuremath{\textsf{E}}} 
\newcommand{\Var}{\ensuremath{\textsf{Var}}} 
\newcommand{\Reals}{\ensuremath{\mathbb{R}}}
\DeclareMathOperator{\di}{d \!}

\newcommand{\sql}[1]{\texttt{#1}}

\newcommand{\join}{\ensuremath{\mathbin{\Join}}}
\newcommand{\semijoin}{\ensuremath{\mathbin{\ltimes}}}
\newcommand{\select}{\ensuremath{\sigma}}

\newcommand{\OurSys}{\textsf{PARQO}}
\newcommand{\plan}{\ensuremath{\pi}}
\newcommand{\PlanSpace}{\ensuremath{\Space\Pi}}
\newcommand{\sel}{\ensuremath{s}}
\newcommand{\sels}{\ensuremath{\mathbf{s}}}
\newcommand{\selerr}{\ensuremath{\varepsilon}}

\newcommand{\Sel}{\ensuremath{S}}
\newcommand{\Sels}{\ensuremath{\mathbf{S}}}

\newcommand{\Cost}{\ensuremath{\narrow{Cost}}}
\newcommand{\Penalty}{\ensuremath{\narrow{Penalty}}}
\newcommand{\optplan}{\ensuremath{\pi^\star}}
\newcommand{\optCost}{\ensuremath{\Cost^\star}}
\newcommand{\Opt}{\ensuremath{\narrow{Opt}}}

\newcommand{\shep}[1]{{\leavevmode\color{black}{#1}}}
\newcommand{\reva}[1]{{\leavevmode\color{black}{#1}}}
\newcommand{\revb}[1]{{\leavevmode\color{black}{#1}}}
\newcommand{\revc}[1]{{\leavevmode\color{black}{#1}}}
\newcommand{\common}[1]{{\leavevmode\color{black}{#1}}}

\newcommand{\sensilocal}{\ensuremath{\xi^\narrow{\,local}}}
\newcommand{\sensimorris}{\ensuremath{\xi^\narrow{\,morris}}}
\newcommand{\sensisobol}{\ensuremath{\xi^\narrow{\,sobol}}}
\newcommand{\EE}{\ensuremath{\narrow{EE}}}
\newcommand{\seq}[2]{\ensuremath{#1}^{\scriptscriptstyle[#2]}}
\newcommand{\KL}{\ensuremath{\narrow{KL}}}

\DeclareMathOperator*{\argmin}{arg\,min}

\usepackage{xcolor}

\definecolor{Magenta}{rgb}{1,0,1}
\definecolor{Plum}{HTML}{800080}

\newcommand\vldbdoi{XX.XX/XXX.XX}
\newcommand\vldbpages{XXX-XXX}
\newcommand\vldbvolume{17}
\newcommand\vldbissue{1}
\newcommand\vldbyear{2024}
\newcommand\vldbauthors{\authors}
\newcommand\vldbtitle{\shorttitle} 
\newcommand\vldbavailabilityurl{URL_TO_YOUR_ARTIFACTS}
\newcommand\vldbpagestyle{plain} 
\begin{document}

\title{\OurSys: Penalty-Aware Robust \shep{Plan Selection in} \mbox{Query Optimization}}


\author{Haibo Xiu, Pankaj K. Agarwal, and Jun Yang}

\affiliation{%
  \institution{Duke University, Durham, NC, USA}
}
\email{haibo.xiu@duke.edu, {pankaj, junyang}@cs.duke.edu}

\begin{abstract}
The effectiveness of a query optimizer relies on the accuracy of selectivity estimates.
The execution plan generated by the optimizer can be extremely poor in reality due to uncertainty in these estimates.
This paper presents \textbf{\OurSys} (\textbf{P}enalty-\textbf{A}ware \textbf{R}obust \shep{Plan Selection in} \textbf{Q}uery \textbf{O}ptimization),
a novel system where users can define powerful robustness metrics
that assess the expected penalty of a plan with respect to true optimal plans
under uncertain selectivity estimates.
\OurSys\ uses workload-informed profiling to build error models,
and employs principled sensitivity analysis techniques to identify human-interpretable selectivity dimensions with the largest impact on penalty.
Experiments on three benchmarks demonstrate that \OurSys\ finds robust, performant plans,
and enables efficient and effective parametric optimization.
\end{abstract}

\maketitle

\pagestyle{\vldbpagestyle}
\begingroup\footnotesize\noindent\raggedright\textbf{PVLDB Reference Format:}\\
\footnotesize{\vldbauthors. \vldbtitle. PVLDB, \vldbvolume(\vldbissue): \vldbpages, \vldbyear.\\
}\href{https://doi.org/\vldbdoi}{\footnotesize doi:\vldbdoi}
\endgroup
\begingroup
\renewcommand\thefootnote{}\footnote{\noindent
This work is licensed under the Creative Commons BY-NC-ND 4.0 International License. Visit \url{https://creativecommons.org/licenses/by-nc-nd/4.0/} to view a copy of this license. For any use beyond those covered by this license, obtain permission by emailing \href{mailto:info@vldb.org}{info@vldb.org}. Copyright is held by the owner/author(s). Publication rights licensed to the VLDB Endowment. \\
\raggedright Proceedings of the VLDB Endowment, Vol. \vldbvolume, No. \vldbissue\ %
ISSN 2150-8097. \\
\href{https://doi.org/\vldbdoi}{doi:\vldbdoi} \\
}\addtocounter{footnote}{-1}\endgroup

\ifdefempty{\vldbavailabilityurl}{}{
\begingroup\footnotesize\noindent\raggedright\textbf{PVLDB Artifact Availability:}\\
The source code, data, and/or other artifacts have been made available at \url{https://github.com/Hap-Hugh/PARQO}.
\endgroup
}

\section{Introduction}
\label{sec:introduction}

Given a query and a set of candidate execution plans,
a standard cost-based query optimizer chooses the plan with the lowest \emph{estimated cost}.
Errors in cost estimates can lead to suboptimal, and sometimes catastrophic, plan choices.
Research has attributed the main source of such errors to inaccurate selectivity estimates~\cite{scheufele1997complexity, job},
which are used by the optimizer to predict the cardinalities of result sets returned by subplans without executing them.
Despite decades of research on improving selectivity and cardinality estimates,
be it using better data summaries~\cite{srivastava2006isomer, hu2022selectivity}, samples~\cite{wu2001applying}, or machine learning models~\cite{kipf2018learned, wang2020we},
the problem remains unresolved.
Since higher accuracy comes at some cost,
such as runtime monitoring and ongoing maintenance,
a database system must strike a balance between the cost and benefit of accurate estimates.
Therefore, coping with uncertainty in selectivity estimates will likely remain a long-term challenge.
At a high level, this paper studies how uncertainties in selectivity estimates affect plan optimality,
and how to find ``robust'' plans that work well despite such uncertainties.

The idea of \emph{robust query optimization} has been around for years~\cite{chu1999least, chaudhuri2010variance, jayant2008identifying}.
\common{%
Despite extensive research that has introduced various notions of robustness and approaches for finding robust plans (see~\Cref{sec:related} for more discussion), wide adoption of these results has yet to happen in practice.
\shep{To enable impact for this line of research, we argue that we must address several challenges in a concerted effort.}

First, there is no ``one-size-fit-all'' when it comes to the notion of robustness.
For some applications, robustness may mean that the chosen plan's cost is insensitive to errors in cardinality estimates.
Some may care instead about how errors affect the cost optimality of the chosen plan:
if, over all likely estimation errors, a plan still remains optimal or nearly optimal among all alternatives, whether its cost is sensitive to any estimate is irrelevant.
Others may prefer more nuanced robustness definitions:
e.g., to meet a service-level agreement,
they may want to ``penalize'' performance degradation over the optimal proportionally, but only if the degradation exceeds a threshold.
\shep{Despite the myriad of application needs,
systems proposed in previous work tend to specialize in one robustness metric
or make design decisions that implicitly encode specific assumptions about robustness;
this limited applicability dissuades adoption.
Instead, we} strive for a general framework and techniques that support flexible and powerful robustness definitions.

Related to this point, the notion of robustness is strongly tied to the degree of uncertainty.
\shep{In many settings, we naturally accumulate or can proactively acquire knowledge on
how errors are distributed given the data and query workload.
Many previous robust optimization approaches do not leverage this knowledge
and consequently make overly conservative choices that are forced by unlikely scenarios and perform poorly in common ones.
Instead, we want a framework that incorporates knowledge about uncertainty in a principled way}
when defining and optimizing robustness.

Second, robust query optimization inherits all scalability and efficiency challenges of traditional query optimization and then adds more.
Real queries often contain many joining tables and predicates;
for example, in the JOB benchmark~\cite{job} based on a real dataset (IMDB),
query Q29 is a 17-way join, with two tables involved in self-joins. PostgreSQL makes more than 13{,}000 cardinality estimates when optimizing this query.
Such high dimensionality of the selectivity space makes the problem of robust query optimization very challenging.
For many useful robustness metrics, it is impossible to assess a plan's robustness by itself,
without examining how other competing plans would perform under uncertainty in high-dimensional space.
Therefore, we must tame the overhead of robust query optimization in order to justify its use.

Third, practical adoption may also require integration with existing database systems.
Past research has seen many examples where
robust query optimization requires modifications to traditional database optimizers \shep{and} execution engines.
Coupled with the limitation that they only support specific notions of robustness,
adoption is a hard sell.
Therefore, there is an argument for solutions that interface with traditional systems
to support more general notions of robustness in a scalable and efficient way.

To address these challenges, we introduce \emph{\textbf{\OurSys}} (\emph{\textbf{P}enalty-\textbf{A}ware \textbf{R}obust \shep{Plan Selection in} \textbf{Q}uery \textbf{O}ptimization}):}
\begin{itemize}[leftmargin=*]
    \item We develop a framework that takes the general approach of \emph{stochastic optimization}~\cite{spall2005introduction}
    and defines the robustness objective as minimizing \emph{expected penalty}.
    The key components in this definition include a flexible user-defined penalty function,
    which assesses the penalty incurred by a plan with respect to the true optimal,
    and a statistical model of the selectivity estimation errors,
    which we show how to obtain by profiling the database workload.
    This powerful combination allows \OurSys\ to tailor to a broader range of application needs.
    To the best of our knowledge, previous work (\Cref{sec:related})
    either does not follow a stochastic optimization approach \shep{to leverage the statistical knowledge about errors},
    or chooses less general robustness measures (and/or employs heuristics reflecting such objectives)
    that depend only on an individual plan (e.g., whether its own cost is sensitive to error)
    but not how it compares with other alternatives (such as the true optimal).

    \item On taming high dimensionality of the optimization problem,
    existing heuristics for selecting dimensions are not always aligned to the definition of robustness,
    and methodically, they rely on rather rudimentary techniques that analyze dimensions one at a time.
    We borrow principled techniques from the \emph{sensitivity analysis} literature~\cite{saltelli2010variance, wainwright2014making}
    that account for interactions among multiple dimensions,
    and adapt them to our setting.
    Our selection of \emph{sensitive dimensions} considers the given expected penalty objective,
    and therefore is fully aware of and tailored to the error distribution as well as the penalty definition.
    Finally, selected sensitive dimensions correspond to selection/join condition combinations in the query,
    which are interpretable and actionable --- for example,
    a user may investigate a sensitive dimension and improve its accuracy \revb{(by selectively reanalyzing statistics and/or sampling)} in order to obtain a better plan.

    \item Given a query and its selectivity estimates, we show how to find a robust plan that minimizes expected penalty,
    focusing on the sensitive dimensions.
    Except for very expensive queries, the amount of work that goes into finding a robust plan from end to end
    (including the identification of sensitive dimensions through sensitivity analysis) may not
    justify doing so to optimize just a single execution.
    We show how to reuse the work in robust query optimization and amortize its cost 
    in the setting of \emph{parametric query optimization} (\emph{PQO})~\cite{ioannidis1997parametric},
    where the optimization overhead is shared among multiple queries with the same template but different query parameters,
    which frequently arise in practice.

    \item \common{Despite the generality of our framework,
    \OurSys\ is designed to work with existing optimizers and cardinality estimation methods.
    We have implemented it on top of PostgreSQL and conducted an end-to-end evaluation using three popular benchmarks: JOB~\cite{job}, DSB~\cite{ding2021dsb}, and STATS-CEB~\cite{han2021cardinality}.}
    We demonstrate how \OurSys\ is able to suggest hints that are interpretable and actionable for understanding and improving query performance,
    how resilient \OurSys's robust plans are against inaccurate selectivity estimates and how they soundly outperform traditional plans,
    and finally, how \OurSys\ delivers significant benefits to multiple queries in the PQO setting.
    We illustrate \OurSys's effectiveness in these scenarios using an example below.
\end{itemize}

\begin{example}\itshape\label{eg:q17-intro}
Consider Q17 below from the JOB benchmark~\cite{job}.
\vspace*{0.5ex}
\begin{lstlisting}[language=SQL, basicstyle=\footnotesize\ttfamily]
SELECT MIN(n.name) AS member_in_charnamed_american_movie,
       MIN(n.name) AS a1
FROM cast_info AS ci, company_name AS cn, keyword AS k, 
     movie_companies AS mc, movie_keyword AS mk, 
     name AS n, title AS t
WHERE cn.country_code ='[us]' AND k.keyword ='character-name-in-title' AND n.name LIKE 'B%'
  AND n.id = ci.person_id AND ci.movie_id = t.id 
  AND t.id = mk.movie_id AND mk.keyword_id = k.id 
  AND t.id = mc.movie_id AND mc.company_id = cn.id
  AND ci.movie_id = mc.movie_id AND ci.movie_id = mk.movie_id
  AND mc.movie_id = mk.movie_id;
\end{lstlisting}
Based on the error profiles on selectivity estimation collected from a workload (not specifically for this query),
\OurSys\ carries out a sensitivity analysis for PostgreSQL's plan for Q17
and identifies the most sensitive selectivity dimension to be $mk \join k^\select$
(the $\select$ superscript on a table indicates the presence of a local selection condition).
This suggestion is interpreted and actionable.
Indeed, if we pay extra diligence to learn the true selectivity of $mk \join k^\select$ and hint it to PostgreSQL,
the new plan will achieve a \common{$5.84\times$} speedup in actual execution time.
Additional details and more experiments can be found in \Cref{sec:expr}.

\OurSys\ can also suggest a \emph{robust plan} for Q17.
To get a sense of how this plan would fare in real-world situations where selectivity errors arise inevitably due to data updates,
we simulated a scenario of an evolving database by time-partitioning the IMDB database used by JOB into 9 different instances (DB1 through DB9),
each with titles and associated data from a contiguous time period.
\OurSys\ only has access to DB5 when choosing the robust plan,
and we execute this same plan on all 9 instances and compare its running time
with PostgreSQL's plans (each obtained for the specific instance).
As can be seen from \Cref{fig:verify-slide},
\OurSys's single robust plan consistently beats PostgreSQL on all 9 instances, with a \common{$3.86\times$} speedup on average.
This is of course just one data point---more experiments can be found in \Cref{sec:expr}.

Finally, improving the performance of just a single execution would not justify the overhead of robust query optimization,
but \OurSys\ shines when combined with PQO, where we share the optimization overhead across many queries with the same template
--- in this case queries that differ from Q17 only in the choice of literals (e.g., \sql{'[us]'}, \sql{'character-name-in-title'}, \sql{'B\%'}).
By caching and reusing the work done on behalf of Q17,
\OurSys\ eliminates the need to call the optimizer for \common{66\%} of the queries with same template.
Furthermore, for these queries, the average speedup over PostgreSQL plans is \common{$7.14\times$},
resulting in an overall improvement of \common{$2.4\times$} for the entire workload.
Again, we refer readers to \Cref{sec:expr} for additional details.
\end{example}


\vspace{-2mm}
\section{\OurSys\ Framework}
\label{sec:framework}

\mparagraph{Preliminaries and Problem Statement}

A \emph{query template} $Q$ is a query where literal values in its expressions are represented by \emph{parameters}.
To optimize a query with template $Q$ and specific parameters values,
a traditional query optimizer considers a space of \emph{(execution) plans} $\PlanSpace(Q)$.
For each plan $\plan \in \PlanSpace(Q)$,
the optimizer uses a set of \emph{selectivities} $\sels = (\sel_1, \ldots, \sel_d) \in [0,1]^d$ relevant to $Q$
to calculate the cardinalities of results returned by various subplans of $\plan$
and in turn to estimate the overall cost of $\plan$, denoted $\Cost(\plan, \sels)$.
The goal of traditional query optimization is to find the \emph{optimal plan} given selectivities $\sels$,
denoted by $\optplan(Q, \sels) = \argmin_{\plan \in \PlanSpace(Q)} \Cost(\plan, \sels)$;
we further denote its cost by $\optCost(Q, \sels)$.
When it is clear that we are referring to a given template $Q$, we omit $Q$ from these notations.

In reality, we do not have the true selectivities $\sels$, but only their estimates instead.
Acting on this uncertain information, suppose the optimizer picks a plan $\plan$.
We would like to quantify the \emph{penalty} incurred by executing $\plan$ relative to the real optimal plan,
where their costs are based on the true selectivities $\sels$.
There are many reasonable options for defining penalty.
For example, it can be defined using a tolerance factor $\tau$:
\begin{equation}\label{eq:penalty}\resizebox{.92\hsize}{!}{$
\begin{split}
  \Penalty(\plan, \sels) =
  \begin{cases}
      0 & \text{ if } \Cost(\plan, \sels) \le (1 + \tau) \cdot \optCost(\sels),\\
      \Cost(\plan, \sels) - \optCost(\sels) & \text{ otherwise.}
  \end{cases}
\end{split}
$}\end{equation}
In other words, the penalty is proportional to the amount of cost exceeding the optimal,
but only if it is beyond the prescribed tolerance.
This particular definition would capture the scenario where
a provider aims to fulfill a service-level agreement under which
any performance degradation above a certain threshold will incur a proportional monetary penalty.
\shep{As motivated in \Cref{sec:introduction},
to make \OurSys\ more broadly applicable,
our framework works with any user-defined penalty definition, not just the above example.
See the extended version of this paper~\cite{fullversion} for other possibilities:
e.g., probability of exceeding the tolerance threshold, standard deviation in cost difference, or simply the cost difference itself, etc.}

Since we do not know true selectivities $\sels$ in advance, we cannot evaluate the penalty directly at optimization time.
Instead, \OurSys\ models selectivities as a random vector $\Sels$,
and evaluates the expected penalty $\Expectation[\Penalty(\plan, \Sels) | \hat\sels]$.
Let $f(\sels | \hat\sels)$ denote the probability density function for the distribution of true selectivities $\sels$ conditioned on the current estimate $\hat\sels$.
We formally define the problem of \emph{finding a robust plan} as follows:
\begin{itemize}[leftmargin=*]\itshape
\item[]\textbf{(Robust plan)}
    Given a query with template $Q$, selectivity estimates $\hat\sels \in [0,1]^d$,
    and a conditional distribution of true selectivities $\Sels \sim f(\sels | \hat\sels)$,
    find a plan $\plan \in \PlanSpace(Q)$ that minimizes:
    \begin{equation}\label{eq:expected-penalty}
        \Expectation[\Penalty(\plan, \Sels) | \hat\sels] = \int \Penalty(\plan, \sels) \cdot f(\sels | \hat\sels) \di \sels.
    \end{equation}    
\end{itemize}

We also define the problem of \emph{finding sensitive (selectivity) dimensions}, informally at this point, as follows:
\begin{itemize}[leftmargin=*]\itshape
\item[]\textbf{(Sensitive selectivity dimensions)}
    Given $Q$, $\hat\sels$, $f(\sels | \hat\sels)$, and a plan $\plan \in \PlanSpace(Q)$,
    find up to $k$ dimensions among $1, \ldots, d$ having the ``largest impact'' on $\Penalty(\plan, \sels)$.
\end{itemize}
We defer a detailed discussion on various options of defining ``largest impact'' to \Cref{sec:sensitivity},
but as a preview, \OurSys\ prefers defining the impact of selectivity dimension $i$
as the contribution to the variance in $\Penalty(\plan, \Sels)$ due to uncertainty in $\sel_i$.

\mparagraph{Remarks}

\shep{As mentioned earlier, our framework works with other penalty definitions,
but we choose the one in \Cref{eq:expected-penalty} for our experiments in \Cref{sec:expr},
because this definition} is easy to interpret yet still illustrates two important features supported by our framework.
First, it is defined relative to the would-be optimal plan,
allowing it to model a broader range of notions of robustness than those that are defined only using the plan $\plan$ itself
(such as how sensitive $\Cost(\plan, \sels)$ is to variation in $\sels$).
Second, it is not merely linear in $\Cost(\plan, \sels)$, which would make the problem considerably easier because of linearity of expectation.%
\footnote{\label{footnote:other-penalties}For example, consider the alternative definition of $\Penalty(\plan, \sels) = \Cost(\plan, \sels) - \optCost(\sels)$.
    Because of the linearity of expectation, the optimization problem boils down to a much simpler version
    of minimizing expected cost $\Expectation[\Cost(\plan, \Sels)]$, which is independent of the optimal plan costs.
    While this definition may be appropriate if our overall goal is system throughput,
    it does not particularly penalize bad cases, which users with low risk tolerance may be more concerned with.
    \revb{As another example that is ``pseudo-dependent'' on the optimal plan costs,
    the \emph{P-error} metric recently proposed in~\cite{han2021cardinality} defines
    $\Penalty(\plan, \sels) = \Cost(\plan, \sels) / \optCost(\sels)$,
    but let us consider the logarithm of P-error instead.
    Because $\log(\Cost(\plan, \sels) / \optCost(\sels)) = \log\Cost(\plan, \sels) - \log\optCost(\sels)$,
    we see that minimizing expected log-P-error again can be done without regard to the optimal plan costs by the linearity of expectation.}}
We want to have a framework and techniques capable of handling more general cases.

Finally, we acknowledge that besides selectivity estimation errors, many other issues also contribute to poor plan quality,
including inaccuracy in the cost function $\Cost(\plan, \sels)$ as well as suboptimality of the optimization algorithm;
we focus only on selectivity estimation because it has been identified as the primary culprit~\cite{job, scheufele1997complexity}.
In the remainder of this paper, we shall assume that $\Cost$ is exact and
that we can obtain the optimal plan $\optplan(\sels)$ if given true selectivities.

\mparagraph{System Overview and Paper Outline}

\OurSys\ is designed to work with any traditional query optimizer that supports (or can be extended to support) two primitives:
$\Opt(Q, \sels)$ returns the optimal plan $\optplan(\sels)$ for $Q$ given selectivities $\sels$;
$\Cost(\plan, \sels)$ returns the cost of plan $\plan$ for selectivities $\sels$.
We followed the strategy of \cite{han2021cardinality} and the \texttt{pg\_hint\_plan} extension~\cite{pg_hint_plan}
to inject $\sels$ and $\plan$ into PostgreSQL for our implementation.
When analyzing the complexity of our algorithms, we count the number of calls to $\Opt$ and $\Cost$.
Note that $\Cost$ is much cheaper than $\Opt$.

A prerequisite of our framework is the distribution $f(\sels | \hat\sels)$ of true selectivities conditioned on their estimates.
While any distribution could be plugged in, including non-informative ones in case no prior knowledge is available,
an informative distribution will make \OurSys\ more effective.
We outline a strategy in \Cref{sec:error} for inferring this distribution
by collecting error profiles for query fragments called \emph{querylets} from the database workload.
These profiles are able to capture \common{some} errors due to dependencies among query predicates.
Finally, \Cref{sec:error} also clarifies what relevant selectivity dimensions are for a given query template.

Next, building on this knowledge of how errors are distributed, \Cref{sec:sensitivity} tackles the problem of finding sensitive dimensions,
for a given query plan $\plan$, obtained under selectivity estimates $\hat\sels$.
We employ principled sensitivity analysis methods to identify a handful of selectivity dimensions
with biggest impact on the user-defined penalty function.
In particular, \OurSys\ proposes using \emph{Sobol's method}~\cite{SOBOL2001271, saltelli2010variance}, which offers an interpretable measure of ``impact''
based on an analysis of the variance in $\Penalty(\plan, \Sels)$ over $\Sels \sim f(\sels | \hat\sels)$.
We also show in \Cref{sec:sensitivity:hints} how automatically identified sensitive dimensions
can help performance debugging of query plans.

\Cref{sec:robust} describes an algorithm for finding robust query plans by focusing on the selectivity subspace consisting of only the sensitive dimensions.
By sampling from the distribution of true selectivities conditioned on their estimates,
we build a pool of candidate robust plans and select the one with the lowest expected penalty.
We show how sample caching and reuse can significantly reduce the number of $\Opt$ and $\Cost$ calls.
To further mitigate the overhead of robust query optimization,
\OurSys\ combines it with parametric query optimization
so that the work devoted to finding a robust plan can be reused for a different query with the same template.
We develop a principled test for determining when to allow such reuse.

Finally, \Cref{sec:expr} presents a full experimental evaluation of \OurSys\ using \common{three different benchmarks};
\Cref{sec:related} discusses related work;
\Cref{sec:conclude} concludes and outlines future directions.

\vspace{-1mm}
\section{Error Profiling}
\label{sec:error}

The goal of this step is to build a model that approximates $f(\sels | \hat\sels)$
given a query with template $Q$ and selectivity estimates $\hat\sels$,
or equivalently, a model of the error between $\Sels$ and $\hat\sels$.
Some learned selectivity estimators are able to output estimates as well as some measures of uncertainty,
which we may readily adopt if we deem them reliable.
However, we still need a procedure for obtaining $f(\sels | \hat\sels)$ in the general case where such measures are not already available.
Despite the notation $f(\sels | \hat\sels)$, which involves the true selectivities $\sels$,
we do not want to supplant the original selectivity model;
instead, we simply seek to characterize the errors.
Nonetheless, there are some high-level desiderata.
First, we would like this model be informed by the database workload.%
\footnote{\label{footnote:cold-start}\revc{If no such workload exists to start with,
one can generate a random query workload aimed at coverage,
or simply adopt an non-informative error model that conservatively assumes that true selectivities can be arbitrary in $[0,1]$,
and then redo the process after a query workload emerges.
\shep{As query and/or data workloads drift, error distributions may drift as well.
When significant drifts are detected, a straightforward approach is to redo error profiling and subsequent analysis and optimization.
More efficient handling of such drifts is an interesting direction of future work; see \Cref{sec:conclude} for more discussion.}}}
Second, the independence assumption made by many traditional optimizers is often blamed for throwing off cardinality estimates;
hence, we need to go further than profiling each selection predicate and join predicate in isolation,
so we can account for the effect of their interactions on estimation errors.
One the other hand, it is impractical to track estimation error for every possible subquery that shows up during query optimization
--- recall from \common{\Cref{sec:introduction}} that PostgreSQL invokes more than 13{,}000 cardinality estimates for optimizing Q29 alone.
Guided by these considerations, \OurSys\ adopts the following design.

\mparagraph{Querylets}
Given a query or query workload, we build one error profile per ``querylet.''
A \emph{querylet} is subquery pattern involving joins and/or local selection conditions,
e.g., $R^\select \join_p S$,
where superscript $\select$ denotes the presence of at least one local selection condition on a table.
Querylets are uniquely identified by the set of tables, join conditions among them, and the subset of the tables with local selection conditions.
During query execution, for each subquery matching a querylet, we track the estimated and actual cardinalities of its result.
We maintain a sample of all such pairs observed for this querylet in a workload, which constitutes its \emph{error profile}.

We cannot afford to profile all possible querylets, so we choose the following:
all single-table querylets,
all two-table querylets,
plus any additional three-table querylet with the pattern $R^\select \join_{p_1} S \join_{p_2} T$,
if it appears in some query where none of $S$ and $T$ has any local selection.
The cutoff at length two to three is for practicality.
The allowance for some three-table querylets is to capture at least some data dependency beyond binary joins.

For example, in Q17 (\Cref{eg:q17-intro}), one querylet would be $n^\select$, which covers all local selection conditions on $n$.
Another example is $mc \join cn^\select$.
A third example would be $k^\select \join mk \join ci$:
since $mk$ and $ci$ have no selection conditions,
this querylet captures any potential dependency between the local selection on $k$ and the join between $mk$ and $ci$.
As an example of a 3-table querylet that is not profiled,
consider $mc^\select \join t \join cn^\select$ (this case does not arise in Q17),
because both $mc^\select \join t$ and $t \join cn^\select$ would have been profiled already.

Note that one could choose to further differentiate querylets by the columns or query constants involved in the selection conditions,
at the expense of collecting more error files.
For this paper, we specifically want to keep error profiling simple and practical, so we did not explore more sophisticated strategies.
Despite this rather coarse level of error profiling, we obtain good results in practice in \Cref{sec:expr}.
That said, there are particular cases where we observe limitation of our current approach (also further explained in \Cref{sec:expr}).
Our framework allows for any error model to be plugged in, so further improvements are certainly possible.

\mparagraph{Relevant Dimensions and Error Distributions}
For a query template $Q$, we derive the set of relevant dimensions and corresponding error distributions from the set of querylets contained in the template.
Specifically, for each table $R$ with local selection in $Q$, we use the querylet $R^\select$ (otherwise the estimate should be precise).
For each join condition in $Q$, say between $R$ and $S$, we select the most specific two-table querylet matching $Q$.
For example, Q15 of JOB joins $mc$ and $cn$ with local selections on both, so the querylet selected is $mc^\select \join cn^\select$.
However, \common{if neither} $R$ and $S$ has any local selection, we look for the most specific three-table querylets we have profiled.
If there are multiple such error files, we simply merge them.
For example, in Q17, neither $\mathit{ci}$ or $\mathit{mc}$ has any local selection,
but two three-table querylets matching Q17 contain $\mathit{ci}$ and $\mathit{mc}$: $n^\select \join \mathit{ci} \join \mathit{mc}$ and $\mathit{ci} \join \mathit{mc} \join \mathit{cn}^\select$.
We merge the collected error data according to these two querylets together
and build one error distribution attributed to the join between $\mathit{ci}$ and $\mathit{mc}$.
In the end, the set of relevant selectivities correspond to the set of selection and join conditions in the query template.

As a complete example, for Q17, we arrive at $d = 12$ \revb{relevant dimensions as follows.
Error profiles for the three local selection selectivities are readily derived from single-table querylets $\mathit{cn}^\sigma$, $k^\sigma$, and $n^\sigma$.
Note that $4$ tables have no local selections in Q17; we do not consider them relevant dimensions because base table cardinalities are not estimated.
Error profiles for three (out of nine) relevant join selectivities are derived from two-table querylets $n^\sigma \join \mathit{ci}$, $\mathit{mc} \join \mathit{cn}^\sigma$, and $\mathit{mk} \join k^\sigma$.
Error profiles for the next three relevant join selectivities, for $t \join ci$, $t \join \mathit{mc}$, and $t \join \mathit{mk}$,
are derived from three-table querylets $t \join \mathit{ci} \join n^\select$, $t \join \mathit{mc} \join \mathit{cn}^\select$, $t \join \mathit{mk} \join k^\sigma$, respectively.
Finally, for $\mathit{mc} \join \mathit{ci}$, we derive its error profile by merging error profiles for three-table querylets 
$\mathit{cn}^\select \join \mathit{mc} \join \mathit{ci}$ and $\mathit{mc} \join \mathit{ci} \join n^\select$;
for $\mathit{mk} \join \mathit{ci}$, we merge 
$k^\select \join \mathit{mk} \join \mathit{ci}$ and $\mathit{mk} \join \mathit{ci} \join n^\select$;
and for $\mathit{mk} \join \mathit{mc}$, we merge
$k^\select \join \mathit{mk} \join \mathit{mc}$ and $\mathit{mk} \join \mathit{mc} \join \mathit{cn}^\select$.}

For each selectivity $\sel_i$, we create two models, one for low selectivity estimates and one for high selectivity estimates.
\revb{In this paper, we set the low-high cutoff as the median error observed in $\sel_i$'s error profile.}
This simple bucketization is motivated by the observation that errors tend to differ across low and high estimates:
e.g., high selectivity estimates naturally have less room for overestimation.
Each model simply uses a kernel density estimator to approximate the distribution of \emph{log-relative} errors calculated from the error profiles.
Given an estimate $\hat \sel_i$, we pick one of the two models to predict its error depending on how $\hat \sel_i$ compares with the low-high cutoff.
We use $g_i(\selerr_i | \hat \sel_i)$ to denote this combined density estimator for log-relative errors in dimension $i$.

Finally, to put together the error distribution in $\hat\sels$ in the full $d$-dimensional selectivity space,
we assume independence of errors \common{estimated by the $g_i$'s.
Therefore, the conditional pdf in \Cref{eq:expected-penalty} is approximated using the following factorized form}:
\begin{equation}\label{eq:sel-pdf}
    f(\sels | \hat\sels) \approx \prod_{i=1}^d g_i(\log(\hat\sel_i / \sel_i) | \hat\sel_i).
\end{equation}

\vspace*{-3ex}
\common{
\mparagraph{Discussion}
It is worth noting that while we assume independence among the $g_i$'s above,
those $g_i$'s derived from the error profiles of two- and three-table querylets already capture dependencies among the join and selection conditions appearing together in them in a query workload.
This approach follows the same intuition as the factor-graph representations for high-dimensional distributions to avoid the high cost of tracking the full distribution.
To demonstrate the effectiveness of this approach, we experimentally validate in \Cref{sec:expr} its advantage
over a baseline where errors for join and selection selectivities are separately and independently profiled.

Of course, since we cap the size of querylets to profile at $3$, dependencies that span longer join chains are not captured.
We also note that our $g_i$'s are rather coarse:
higher accuracy can certainly be achieved by higher-resolution models and additional profiling effort,
e.g., with finer-grained buckets and separate models for different forms of predicates.
More sophisticated models can be easily plugged in;
\OurSys\ only assumes that we can efficiently draw samples from the error distribution.
Here, we only wish to demonstrate a simple approach that does a reasonable job;
our overall model size is under 15KB for each of the three benchmarks tested in \Cref{sec:expr}.}

\mparagraph{A Note on Recentering}
A good estimator should not exhibit a large bias,
meaning that its error distribution should have a mean around $0$.
After error profiling for PostgreSQL, however, we have observed that this is sometimes not the case.
Since \OurSys\ uses error profiles, it is fair to ask how much of its overall advantage simply comes from more careful modeling of errors.
To this end, in \Cref{sec:expr}, we also experimented with a simple fix called \emph{recentering},
where we calculate the expectation of the true selectivities based on $f(\sels | \hat\sels)$
and ask PostgreSQL to use them in optimization.
As we shall see in \Cref{sec:expr}, while this simple fix shows some improvements,
\OurSys\ overall is able to achieve much more.

\section{Sensitivity Analysis}
\label{sec:sensitivity}

Given a query template $Q$ and selectivity estimates $\hat\sels \in [0,1]^d$,
consider the plan $\plan$ chosen by a traditional optimizer $\hat\sels$: i.e., $\plan = \optplan(\hat \sels)$.
Given $f(\sels | \hat\sels)$, we want to select a subset of up to $k$ out of $d$ dimensions as \emph{sensitive dimensions}.
We have two goals.
First, we would like these dimensions to serve as interpretable and actionable hints that help user understand and improve the performance of $\plan$.
Second, for the subsequent task of finding robust plans, we would like sensitive dimensions to help us reduce dimensionality and tame complexity.
In the following, we will first review previous approaches and basic sensitivity analysis methods,
and then introduce more principled methods.
Then, we briefly discuss how sensitive dimensions can be used to help tune query performance.
\vspace{-1.5mm}
\subsection{From Local to Global Analysis}
\label{sec:sensitivity:local-global}

Before presenting \OurSys's approach, we first briefly explain some alternative approaches for contrast.
Given a plan $\plan$, a number of previous papers~\cite{purandaredimensionality, 2018rqo, jayant2008identifying} define the sensitivity of a dimension $i$
using merely the local properties of the plan's cost function, e.g.,
the partial derivative $\partial\,\Cost(\plan, \sels) / \partial\sel_i$ respect to dimension $i$
evaluated at $\hat\sels$, the current selectivity estimates.
One fundamental limitation of this definition is that it does not address the question ``what would we have done differently.''
It may well be the case that the cost of $\plan$ is highly sensitive to $\sel_i$,
but the optimality of $\plan$ (or its penalty with respect to the optimal plan) is insensitive to $\sel_i$ for all likely values of $\sel_i$.
Hence, \OurSys\ focuses instead on \common{penalty-aware} analysis.

One obvious improvement is to replace the cost function with the penalty function, which gives us
$\partial\,\Penalty(\plan, \hat\sels) / \partial\hat\sel_i$ as a penalty-aware sensitivity measure for dimension $i$.
We can further improve it by incorporating our knowledge of the error distribution
and considering the expected penalty incurred by error in each dimension,
resulting in the following definition:
$\sensilocal_i(\plan, \hat\sels) = \Expectation[\Penalty(\plan, \Sels)) | \hat\sels]$,
where $\Sels = (\ldots, \hat\sel_{i-1}, \Sel_i, \hat\sel_{i+1}, \ldots)$ have identical component values as $\hat\sels$ except dimension $i$
for which $\Sel_i \sim g_i(\log(\hat\sel_i / \sel_i) | \hat\sel_i)$ (see also \Cref{eq:sel-pdf}).
However, such a definition is still limited to \emph{One-At-a-Time} (\emph{OAT}) analysis,
which fails to capture interaction among errors across dimensions.
In the following, we present principled methods for \emph{global} sensitive analysis to overcome this limitation.

\newcommand{\sample}{\ensuremath{\mathbf{x}}}
\newcommand{\Sample}{\ensuremath{\mathbf{X}}}

\reva{%
Two popular methods from the sensitivity analysis literature~\cite{saltelli2010variance, wainwright2014making} are \emph{Morris} and \emph{Sobol's.}
The \emph{Morris Method}~\cite{morris1991factorial} is global in the sense that it considers a collection of ``seeds'' from the whole input space,
but it still relies on local, derivative-based measures (called ``elementary effects'') at each seed that are OAT.
We have adapted this method to our setting to incorporate knowledge of the error distribution; see~\cite{fullversion} for details.
However, as we will see in \Cref{sec:expr}, \emph{Sobol's Method} turns out to be more effective;
therefore, it will be our focus in the following.}

\mparagraph{Sobol's Method}
%
\emph{Sobol's Method}~\cite{SOBOL2001271, saltelli2010variance}, based on analysis of variance,
performs a fully global analysis and accounts for interactions among all dimensions.
Given a function $h: [0,1]^d \to \Reals$, this method considers its stochastic version $Y = h(\Sample)$,
where $\Sample$ is a random input vector characterized by pdf $f_\Sample$.
The variance of $Y$ can be decomposed as follows:
\begin{equation*}\footnotesize
    \Var[Y] = \sum_{1 \le i \le d} V_i
            + \sum_{1 \le i < j \le d} V_{ij}
            + \sum_{1 \le i < j < k \le d} V_{ijk}
            + \cdots
            + V_{1...d}.
\end{equation*}
In the above, each $V_\mathbf{u}$, where $\mathbf{u}$ is a non-empty subset of the dimensions,
is the contribution to the total variance attributed to the interactions among the components of $\mathbf{u}$.
For each input dimension $i$,
$V_i = \Var[\Expectation[Y | X_i]]$,
where the (inner) expectation, conditioned on a particular value for dimension $i$, is over all variations in other dimensions,
and the (outer) variance is over all variations in dimension $i$.
For each subset of two dimensions $i$ and $j$,
$V_{ij} = \Var[\Expectation[Y | X_i X_j]] - V_i - V_j$,
and similarly for larger subsets of dimensions.
Normalizing each $V_\mathbf{u}$ by $\Var[Y]$ yields
the \emph{Sobol's index} $S_\mathbf{u} = V_\mathbf{u}/\Var[Y]$ for the combination of input dimensions in $\mathbf{u}$.
Of particular interests are the so-called
\emph{first-order index} $S_i = V_i / \Var[Y]$, which is the portion of the total variance attributed to $X_i$ alone;
and the \emph{total-order index} $S^T_i = \sum_{i \in \mathbf{u} \subset [1..n]} S_\mathbf{u}$, which is the portion of the total variance that $X_i$ contributes to (alone or together with other dimensions).
The latter can be computed as $S^T_i = 1 - V^T_i / \Var[Y]$, where $V^T_i = \Var[\Expectation[Y | X_{\sim i}]]$, without summing an exponential number of Sobol's indices.

\newcommand{\sampleA}{\ensuremath{\mathbf{a}}}
\newcommand{\sampleB}{\ensuremath{\mathbf{b}}}
\newcommand{\sampleAB}{\ensuremath{\mathbf{a_b}}}
Sobol's indices are computed using a quasi-Monte Carlo method, using $2K$ sample points drawn randomly from $f_\Sample$.
Given two sample points $\sampleA$ and $\sampleB$,
it generates $d$ more points, one for each dimension, by replacing the $i$-th component of $\sampleA$ with the corresponding one in $\sampleB$, obtaining a new point $\seq\sampleAB{i}$.
Given sample points $\sampleA_1, \ldots, \sampleA_K$ and $\sampleB_1, \ldots, \sampleB_K$,
the first-order and total-order indices for dimension $i$ can be estimated through
$V_i \approx \smash{1 \over K} \sum_{j=1}^K h(\sampleB_j) (h(\seq{\sampleAB_j}{i}) - h(\sampleA_j))$
and $V^T_i \approx \smash{1 \over 2K} \sum_{j=1}^K (h(\seq{\sampleAB_j}{i}) - h(\sampleA_j))^2$.

Sobol's method suits our setting perfectly.
Given a plan $\plan$ obtained under selectivity estimates $\hat \sels$,
we analyze the function $h(\sels) = \Penalty(\plan, \sels)$ by drawing the $2K$ samples from $f(\sels | \hat \sels)$.
The first-order and total-order indices give principled and interpretable measures of sensitivities
that are tailored to the user-defined notion of penalty
and are informed by error profiles observed from the database workload.
There are good arguments for using either first-order or total-order indices (or even both);
our current implementation simply uses the first-order indices.

We denote the \emph{Sobol-sensitivity} for dimension $i$ as $\sensisobol_i(\plan, \hat \sels)$.
Overall, this analysis uses $K$ pairs of sample points, each requiring evaluating $\Penalty$ $d+2$ times.
The total cost of Sobol is $O(Kd)$ $\Opt$ and $\Cost$ calls.
We show practical $K$ values to reach convergence in \Cref{sec:expr};
Sobol is generally slower to converge than Morris.
\vspace{-2mm}
\subsection{Sensitive Dimensions as Tuning Hints}
\label{sec:sensitivity:hints}

\OurSys\ uses Sobol-sensitivity by default to identify sensitive selectivity dimensions for a given plan.
Practically, as we have found through experiments in \Cref{sec:expr},
the actual Sobol-sensitivity values of the dimensions make it easy to identify a small number of dimensions that clearly stand out.
\common{For example, for all queries in JOB, this number varies between $2$ to $6$.}
We now describe how these sensitive dimensions are presented by \OurSys\ to users to help them understand and fine-tune plan performance.

Recall from \Cref{sec:error} that all relevant dimensions are pegged to selection and join conditions in the query,
but their error profiles in fact capture more than a single predicate.
Hence, \OurSys\ is careful in presenting such dimensions to users.
For example, the most sensitive dimension for Q17 is $\mathit{mk} \join k^\select$ (\Cref{eg:q17-intro}).
This selectivity needs to be understood as the join selectivity between $\mathit{mk}$ and $k$ \emph{assuming a local selection on $k$},
which is different from the ``plain'' join selectivity of $\mathit{mk} \join k$
(which should have no estimation error at all by itself since it is a join between foreign and primary keys).
The second, and the only other sensitive dimension for Q17, is associated with the join between $\mathit{mk}$ and $ci$,
and will be presented to users as $(\mathit{mk} \semijoin k^\select) \join (\mathit{ci} \semijoin n^\select)$.
Since neither $\mathit{mk}$ nor $\mathit{ci}$ has any local selection in Q17,
the error distribution is derived from the error profiles for querylets
$k^\select \join \mathit{mk} \join \mathit{ci}$ and $\mathit{mk} \join \mathit{ci} \join n^\select$.

With this information, users may decide to investigate further and take action in several ways,
focusing now on these two dimensions instead of all 12 relevant dimensions originally in Q17.
For example, they may want to devote more resources to collecting statistics and/or training models relevant to these two dimensions,
or simply do some additional probing to get better selectivity estimates for these dimensions
and ask the optimizer to re-optimize under these new estimates.
\Cref{eg:q17-intro} already mentioned that correcting the error in the most sensitive dimension ($\mathit{mk} \join k^\select$) leads to a \common{$5.84\times$} speedup in actual execution time of Q17.
If we instead correct the error for the second most sensitive dimension $(\mathit{mk} \semijoin k^\select) \join (\mathit{ci} \semijoin n^\select)$ alone,
the speedup will be \common{$1.38\times$}.
Finally, if we correct both errors, the speedup will be \common{$6.4\times$}.

Beside presenting the sensitive dimensions appropriately to users and allows them to experiment with different selectivities,
\OurSys\ currently does not offer any additional user-friendly interfaces.
There are abundant opportunities for developing future work and applying complementary work (e.g.,~\cite{picaso, jayant2008identifying, tan2022mocha, edu}) on visualizations and interfaces,
such as tools for interactively exploring the penalty and optimal plan landscapes along sensitive dimensions.

\section{Finding Robust Plans}
\label{sec:robust}

Given a query template $Q$ and selectivity estimates $\hat\sels$,
our goal is to find a plan $\plan$ that minimizes the expected penalty $\Expectation[\Penalty(\plan, \Sels) | \hat\sels]$.
This penalty-aware formulation allows for powerful notions of robustness
that are based on global properties of the plan space (since penalties are relative to optimal plans with true selectivities),
as opposed to simple measures such as those based on the local properties of the cost function for $\plan$ itself~\cite{2018rqo}.
This stochastic optimization formulation further enables optimization informed by distributions of selectivity estimation errors observed in workloads,
which are more focused and less conservative than formulations that consider the entire selectivity space, e.g.~\cite{abhirama2010stability, chaudhuri2010variance}.
In the following, we first describe the end-to-end procedure for finding a robust plan for a single query,
and then discuss how to reuse its effort across multiple queries, in the setting of parametric query optimization.
\vspace{-1mm}
\subsection{Finding One Robust Plan}
\label{sec:robust:single}

As the first step, \OurSys\ performs the sensitivity analysis in \Cref{sec:sensitivity}
on the optimizer plan $\optplan(\hat\sels)$ to identify a small subset of sensitive dimensions.
Subsequent steps then operate in the subspace consisting of only the sensitive dimensions.
In remainder of this subsection, with an abuse of notation,
we shall continue to use $d$ for the now reduced number of dimensions
and $\sels, \hat\sels$ for their projected versions;
$f(\sels; \hat\sels)$ would be obtained by \Cref{eq:sel-pdf} in \Cref{sec:error} using only $g_i$'s for sensitive dimensions.

Next, \OurSys\ computes a set of plans, called the \emph{robust candidate plan pool}, as follows.
We draw a sequence of $S$ samples from $f(\sels; \hat\sels)$.
For each sample $\sels$, we call $\Opt$ with these selectivities to obtain the optimal plan $\optplan(\sels)$ at $\sels$ and its cost $\optCost(\sels)$ at $\sels$.
We cache the triple $\langle \sels, \optplan(\sels), \optCost(\sels)\rangle$ (whose purpose will become apparent later),
and register each unique optimal plan in the pool.

Finally, in the third step, for all unique plans in the candidate pool,
\OurSys\ estimates their expected penalties and returns the one with the lowest expected penalty.
Note that this estimation is done using cache populated in the previous step,
since its entries were sampled from $f(\sels; \hat\sels)$ in the first place.
Specifically, we estimate $\Expectation[\Penalty(\plan, \Sels) | \hat\sels]$ as
$\smash{1 \over S} \sum_{\text{cached }\langle \sels^\star, \_, c^\star \rangle} \Penalty(\plan, \sels^\star)$,
where each $\Penalty(\plan, \sels)$ is evaluated using cached $c^\star$ plus a call for $\Cost(\plan, \sels^\star)$.

Overall, not including sensitivity analysis (whose complexity was given in \Cref{sec:sensitivity}),
the process takes $O(S)$ calls to $\Opt$ and $O(S \times \grave S)$ to $\Cost$, where $\grave S$ denotes the number of unique candidate plans.
We show the practical $S$ and $\grave S$ values we used for the JOB benchmark in~\Cref{sec:expr}.
Finally, note that opportunities also exist for caching and reusing the samples acquired during sensitivity analysis.%
\footnote{Strictly speaking, there is a slight difference in their distributions:
    samples in Morris and Sobol (\Cref{sec:sensitivity}) were drawn from the original $f(\sels; \hat\sels)$ with all dimensions,
    whereas samples in this subsection are drawn from $f(\sels; \hat\sels)$ restricted to only the sensitive dimensions.
    This difference can be corrected if needed.}
We did not explore these opportunities in this paper because
we did not want to introduce extra dependencies across components
that may complicate understanding of performance.
\vspace{-1mm}
\subsection{Parametric Robust Query Optimization}
\label{sec:robust:pqo}

PQO works by caching several plans for the same query template as candidates.
Given an incoming query with the same template,
PQO would select one of the cached candidates instead of invoking the optimizer, which is far more expensive.
It is natural for \OurSys\ to combine robust query optimization and PQO,
not only because PQO helps amortize the overhead of robust query optimization across multiple queries,
but also because robust query optimization involves significant effort beyond optimizing for a single point in the selectivity space,
which intuitively should help PQO as well.
This combination allows \OurSys\ to both reduce the optimization overhead and deliver better plans than a traditional optimizer.

Suppose that \OurSys\ has already done the work of optimizing a query with estimated selectivity $\hat\sels$.
Now consider an incoming query with the same template but different parameters and hence different estimated selectivity $\hat\sels'$.
Many opportunities exist to reuse earlier work:
we could assume the same set of sensitive dimensions;
we could reuse the cached optimal plans and costs collected while finding the most robust plan for $\hat\sels$ (\Cref{sec:robust:single});
or we could go as far as returning the same robust plan.
While the last option is the cheapest,
it would either require a stringent reuse condition that limits its applicability,
or give up any form of guarantee on the actual robustness under the new setting.
Hence, \OurSys\ takes a more measured approach, as described below.

First, it would be unrealistic to assume that set of sensitive dimensions always stays the same.
Recall from \Cref{sec:sensitivity} that sensitivity analysis is done for an optimizer plan at a particular setting of selectivity estimates.
We can only expect sensitivity analysis to yield same or similar results
if the penalty ``landscapes'' around $\hat\sels$ and $\hat\sels'$, induced by estimation error, are similar.
We use the KL-divergence between the distributions $f(\sels | \hat\sels)$ and $f(\sels | \hat\sels')$,
denoted $\KL(f(\sels | \hat\sels) \parallel f(\sels | \hat\sels'))$ as a test.%
\footnote{\label{footnote:kl}These distributions are conditioned on the estimates;
    even if the error profiles relative to $\hat\sels$ and $\hat\sels'$ are the same,
    the distributions of true selectivities will have little in common if $\hat\sels$ and $\hat\sels'$ are far away.}
(Importantly, these distributions include \emph{all} dimensions, not merely the sensitive dimensions selected for $\hat\sels$.)
If the KL-divergence is low (we will discuss how to set this threshold shortly),
we allow the set of sensitive dimensions for $\hat\sels$ to be reused for $\hat\sels'$ and continue with other reuse opportunities.
Otherwise, we look for an different $\hat\sels$ to reuse, analyze/optimize $\hat\sels'$ from scratch, or simply fall back to the traditional optimizer.
We argue that the KL-divergence between distributions of true selectivities (conditioned on the estimates)
is a more principled and effective reuse test than those based on surrogates such as similarity among query parameter values.

Now, assuming $\hat\sels'$ has passed the KL-divergence test for reusing $\hat\sels$,
we reuse the $S$ cached samples and the $\grave S$ candidate plans when we optimized for $\hat\sels$.
One complication is that the cached samples were drawn from $f(\sels | \hat\sels)$ instead of $f(\sels | \hat\sels')$.
Hence, when computing expected penalty for a candidate plan $\plan$ at the $\hat\sels'$,
we apply \emph{importance sampling}~\cite{imptsample},
which lets us evaluate properties of a target distribution using samples drawn from a different distribution.
Specifically, the expected penalty of candidate plan $\plan$ can be estimated as:
$\smash{1 \over S}
\sum_{\text{cached }\langle \sels^\star, \_, c^\star \rangle}
\smash{f(\sels^\star | \hat\sels') \over f(\sels^\star | \hat\sels)}
\Penalty(\plan, \sels^\star)$,
where the fraction $\smash{f(\sels^\star | \hat\sels') \over f(\sels^\star | \hat\sels)}$ reweighs the sample
to account for the difference between distributions.
Among the $\grave S$ candidates, we then pick the one with the lowest expected penalty.
With this technique, no $\Opt$ or $\Cost$ calls are needed to find the robust plan for $\hat\sels'$.

If the two distributions are very different, however, importance sampling will require more samples to provide a reasonable estimate.
The lower bound of the sample size required to ensure estimation accuracy through importance sampling is discussed in detail in~\cite{2018samplesize}.
This lower bound is indeed determined by the KL-divergence between the two distributions.
According to this lower bound, we derive the maximum KL-divergence
under which $S$ samples are able to provide acceptable accuracy.
We use this threshold for the reuse test described earlier in this subsection,
ensuring that it is safe to also reuse the same samples for expected penalty calculation.

\vspace{-1mm}
\section{Experiments}
\label{sec:expr}

\revb{We have implemented \OurSys\ on top of \shep{PostgreSQL V16.2}.
We modified PostgreSQL to expose \Opt\ and \Cost\ calls,
with no changes to its optimizer or executor otherwise;
plan and selectivity injection is done as hints to PostgreSQL, with help of \cite{han2021cardinality} and \cite{pg_hint_plan}.
We have open-sourced our implementations in~\cite{PARQO}.}

\common{We use three benchmarks in evaluation.
\textbf{JOB} (Join Order Benchmark)~\cite{job} contains 33 query templates and 113 query instances, with real-world data from IMDB.
This benchmark includes skewed and correlated data distributions as well as and diverse join relationships,
all of which contribute to selectivity estimation errors.
\textbf{DSB}~\cite{ding2021dsb} is an industrial benchmark that builds upon TPC-DS~\cite{nambiar2006making}
by incorporating complex data distributions and join predicates.
\textbf{STATS}(-CEB)~\cite{han2021cardinality} features a real-world dataset from the Stats Stack Exchange.
This paper will focus on evaluation results on JOB, and only summarize the results on DSB and STATS;
additional details are available in~\cite{fullversion}.}
\reva{Each experiment setup involves two query workloads.
First, a \emph{profile workload} is used by \OurSys\ to build error profiles as described in \Cref{sec:error};
example result error distributions can be found in~\cite{fullversion}.
Second, a separate \emph{evaluation workload} contains queries that are targets of our evaluation.
We will describe these workloads when discussing specific experimental setups.}

Unless otherwise specified,
\OurSys\ uses the penalty function in \Cref{eq:penalty} with $\tau = 1.2$,
a setting that is widely used in the robust query optimization literature, e.g., by~\cite{harish2007production, 2018rqo, dutt2014plan, dey2008efficiently}.
\common{When using Morris and Sobol for sensitivity analysis (\shep{\Cref{sec:sensitivity:hints}}),
we sample until convergence, so the parameter $K$ varies across queries;
the number of sensitive dimensions depends on the distribution of scores and also varies.}
To find robust plans (\Cref{sec:robust:single}), we use $S = 100$ samples to build the candidate plan pool for each $Q$;
the number of unique candidate plans per query varies.
\common{We report summary statistics on these varying quantities \shep{later} in \Cref{tab:performance_metrics},
along with other useful measures such as the memory footprint of the error models, the number of relevant dimensions per query, etc.}

All experiments were performed on a Linux server with 16 Intel(R) Core(TM) i9-11900 @ 2.50GHz processors. 
\reva{To reduce noise when measuring execution time, we execute each plan multiple (no fewer than $5$ and up to $101$) times and record the median latency.}

\mparagraph{Traditional vs.\ Robust Plans on Current Database Instance}
As a warm-up, consider a setup where given each query, we compare the actual execution times for the following plans:
\emph{PostgreSQL} denotes the plan found by the PostgreSQL optimizer with its default selectivity estimates $\hat\sels$ (after refreshing all statistics on the current database instance);
\emph{WBM} denotes the plan obtained using the approach of~\cite{2018rqo}%
\footnote{~\cite{2018rqo} proposed three robustness metrics; we show only the plan with the fastest execution time.
WBM sets a threshold (120\%) relative to the cost of the optimizer plan at $\hat\sels$;
it would not consider robust plans with cost higher than this threshold.
};
\emph{Recentering} refers to the baseline introduced in \Cref{sec:error}, where we correct PostgreSQL's estimates using the expectation of $f(\sels | \hat\sels)$ derived from the error profiles;
\emph{\OurSys-Morris} and \emph{\OurSys-Sobol} refer to the plans chosen by \OurSys\ for $\hat\sels$, using the Morris and Sobol's Methods for picking sensitive dimensions, respectively.
\common{Before proceeding, we note that this setup is not ideal for evaluating robust plans.
The advantage of robust plans should be their overall performance over a range of possibilities,
but the current database instance only reflects one of these possibilities.
Nonetheless, given a benchmark database, users inevitably wonder how different plans perform on it, so this setup is natural.
Instead of fixating on one particular query's performance, however, we can get better insight on robustness with an overall comparison over all queries in the workload.
We also will follow up with additional experiments later to examine each plan's performance under different errors and different database instances.}

\begin{figure*}[t]
    \centering
    \begin{subfigure}{0.57\linewidth}
    
    \includegraphics[scale=0.15]{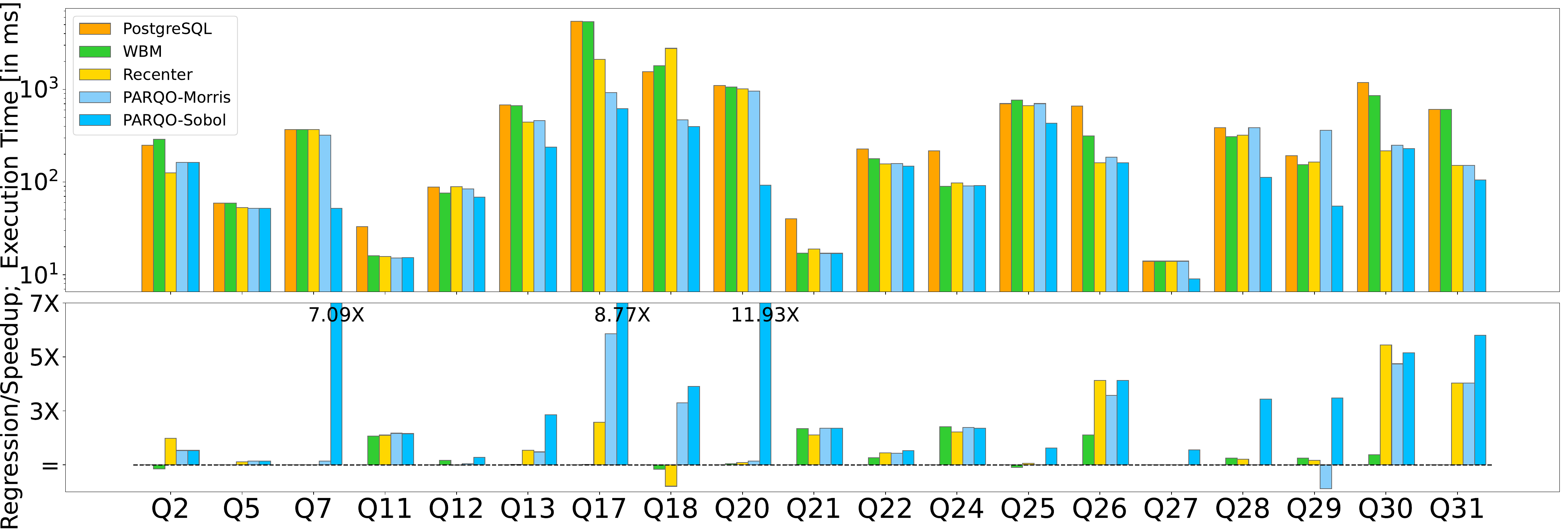}
    \vspace{-6mm}
    \caption{Speedup (19 out of 33)}
    \label{fig:exe-good}
    \end{subfigure}
    \vspace{-2mm}
    \begin{subfigure}{0.37\linewidth}
        \centering
        \includegraphics[scale=0.15]{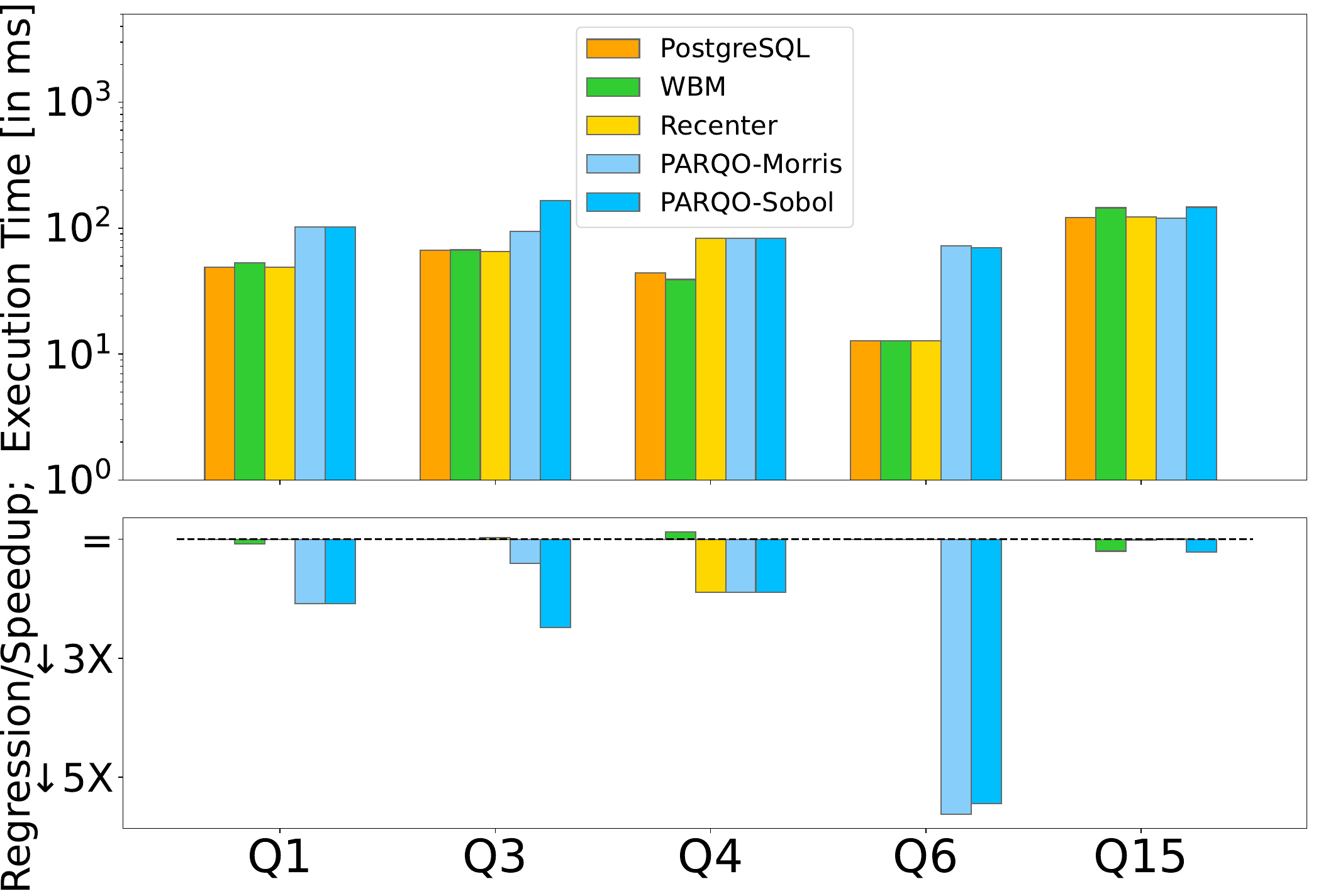}
        \vspace{-2mm}
        \caption{Regression (5 out of 33)}
        \label{fig:exe-bad}
    \end{subfigure}
    \vspace{-2mm}
    \caption{\common{\small{Actual execution times for different plans selected by various methods on JOB.
    (a) and (b) separate queries for which \OurSys-Sobol outperforms or underperforms PostgreSQL;
    queries for which they perform similarly (9 out of 33) are omitted.
    }}}
    \label{fig:exe-all}
    \vspace{-4mm}
\end{figure*}

\Cref{fig:exe-all} summarizes the results on JOB\footnote{\common{In JOB, we use the 33 instances labeled ``(a)'' (one for each query template) as the evaluation workload and all other 80 instances as the profile workload.}}:
the $x$-axis is labeled by queries;
on top we show execution times on a log-scale $y$-axis;
on bottom we additionally show speedup/regression factors on the $y$-axis.
Among the 33 queries in the evaluation workload,
\OurSys-Sobol outperforms PostgreSQL in \common{19} of them (with an overall speedup%
\footnote{Note here and after that we calculate the ``overall'' speedup/regression for a collection of queries as the speedup/regression in the \emph{total execution time} over all queries (as opposed to the arithmetic mean of the speedup/regression factors of individual queries), so speedup/regression in slower queries contribute more than faster queries.}%
\common{ of $4.51\times$}) but underperformed in \common{5} of them.
For clarity, we group them into (a) and (b) in \Cref{fig:exe-all}.
The remaining \common{9} queries are omitted because PostgreSQL and \OurSys-Sobol plans have nearly identical times, and WBM is no faster either.

From \Cref{fig:exe-all}, we see that \OurSys-Sobol outperforms others in most cases.
The most notable improvements are in Q17, where \OurSys-Sobol takes \common{620ms while PostgreSQL and WBM take more than 5{,}000ms,
and in Q20, where \OurSys-Sobol achieves a speedup of $12\times$ over PostgreSQL.}
\OurSys-Morris, although not as effective as \OurSys-Sobol, still surpasses WBM in most cases.

WBM fails to offer much improvement over PostgreSQL here, because it by design avoids plans that cost much higher than PostgreSQL for the original estimates $\hat\sels$,
but this cutoff overlooks the (sometimes likely) possibility that $\hat\sels$ is far off from reality.
As an example, for Q17,
the PostgreSQL plan costs 4.6k (in PostgreSQL cost unit) at $\hat\sels$ while \OurSys-Sobol's plan costs 12.5k;
therefore, WBM did not consider \OurSys-Sobol's plan at all, but instead picked a plan similar to PostgreSQL.
However, it turns out $\hat\sels$ is really off:
in reality, PostgreSQL and WBM ran more than an order of magnitude slower than their cost predicted at $\hat\sels$,
while \OurSys-Sobol ran $>8\times$ faster than them.
This example highlights the need to consider errors instead of relying purely on decisions local to $\hat\sels$.

As for Recentering, it sometimes provides impressive speedups \common{(e.g., Q2, 26, and 30)},
which indicates that our error profiling, despite its simplicity, can already deliver some benefits by correcting biased estimates.
However, Recentering is still far less effective than \OurSys-Sobol overall \common{(e.g., Q7, 18, 20, and more)},
which is evidence that bias correction along is not sufficient --- other components of \OurSys\ also play a significant part in its overall effectiveness.

\begin{figure*}[t]
    
    \begin{minipage}[right]{0.6\linewidth}
    \centering
        \includegraphics[scale=0.175]{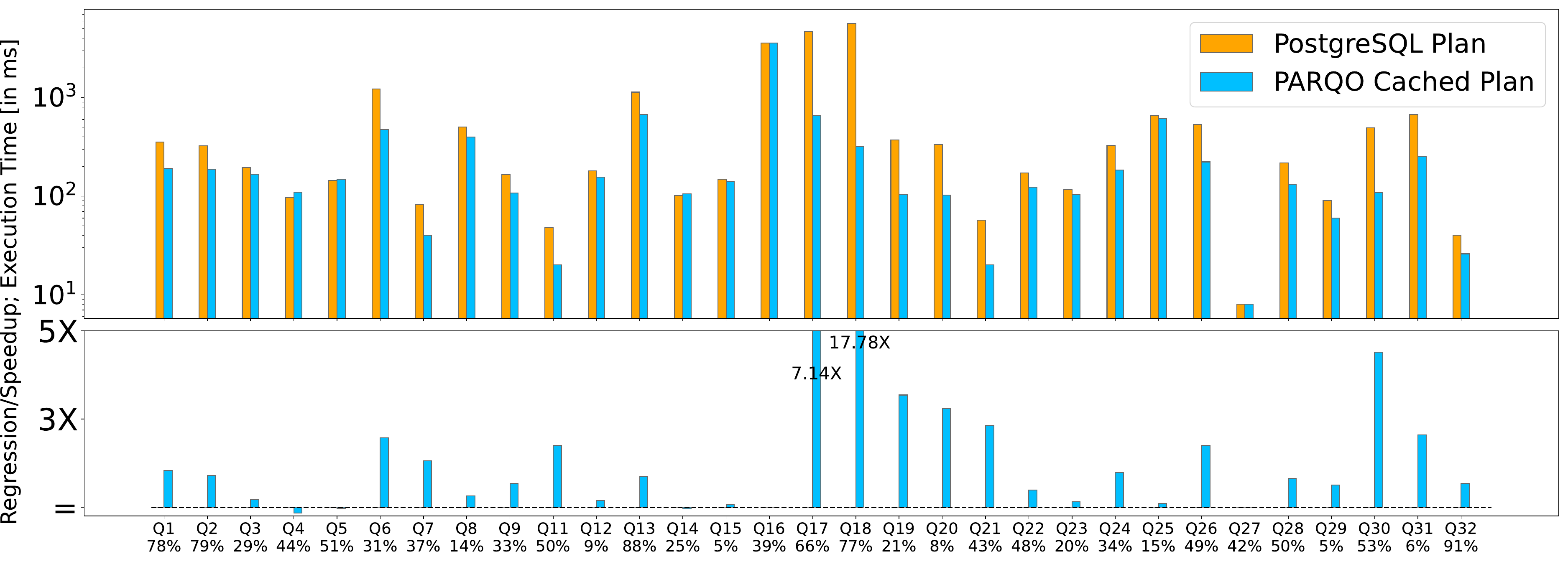}
        \vspace{-8.5mm}
        \caption{\small{\revb{PQO results of JOB. The x-axis shows the template ID and the average reuse fraction of each template.}}}
        \label{fig:pqo}
    \vspace{-3.5mm}
    \end{minipage}
    \begin{minipage}[left]{0.37\linewidth}
    \raggedleft
        
    \vspace{-0.5mm}
    \includegraphics[scale=0.15]{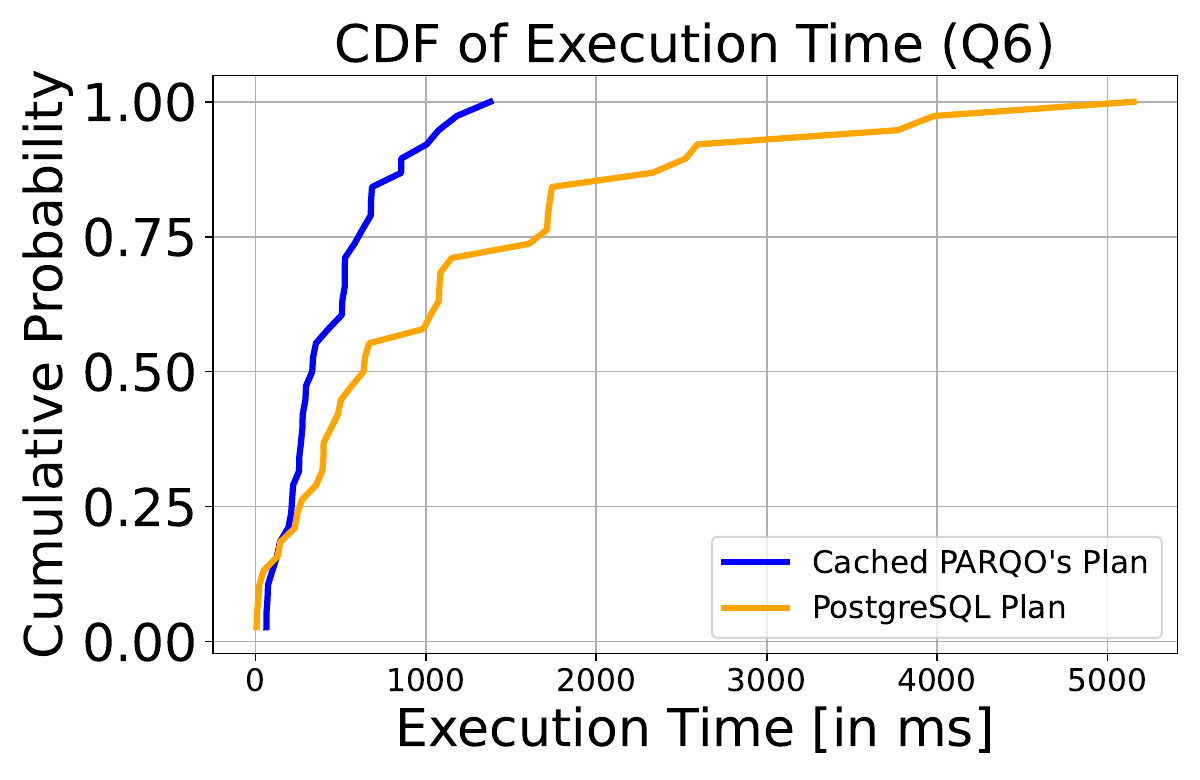}
    \includegraphics[scale=0.15]{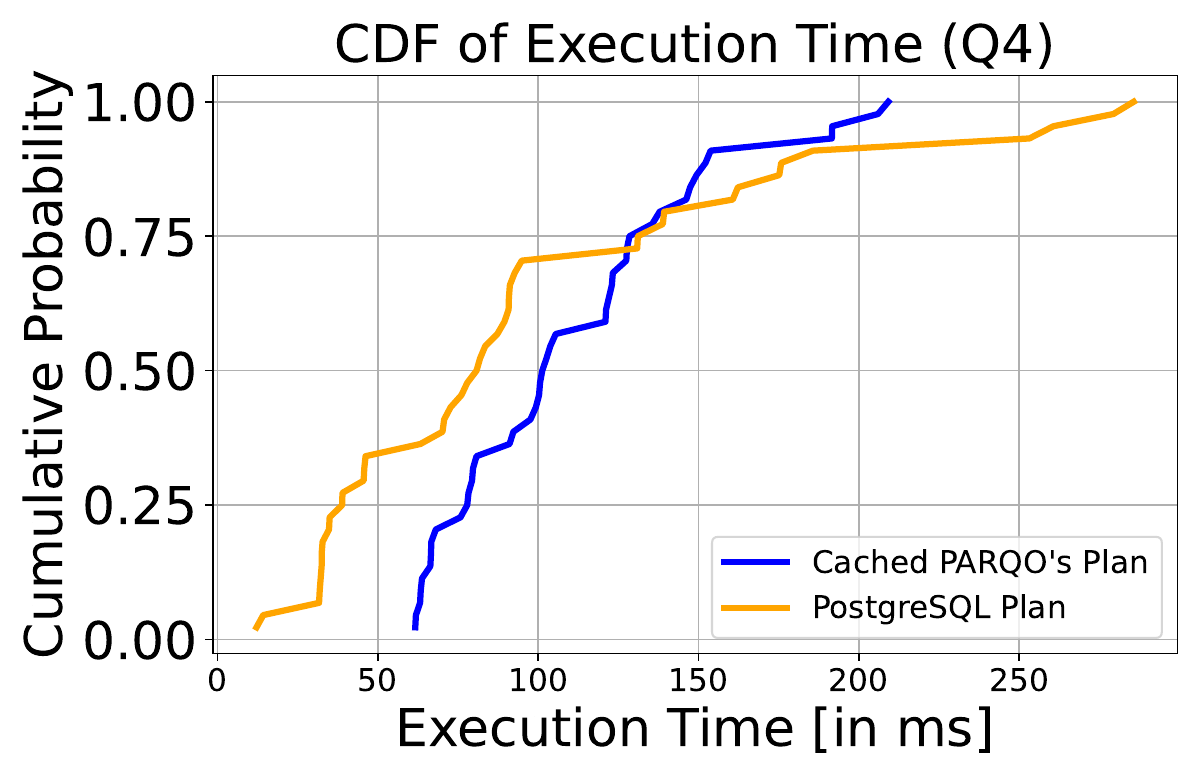}
    \\
    \includegraphics[scale=0.15]{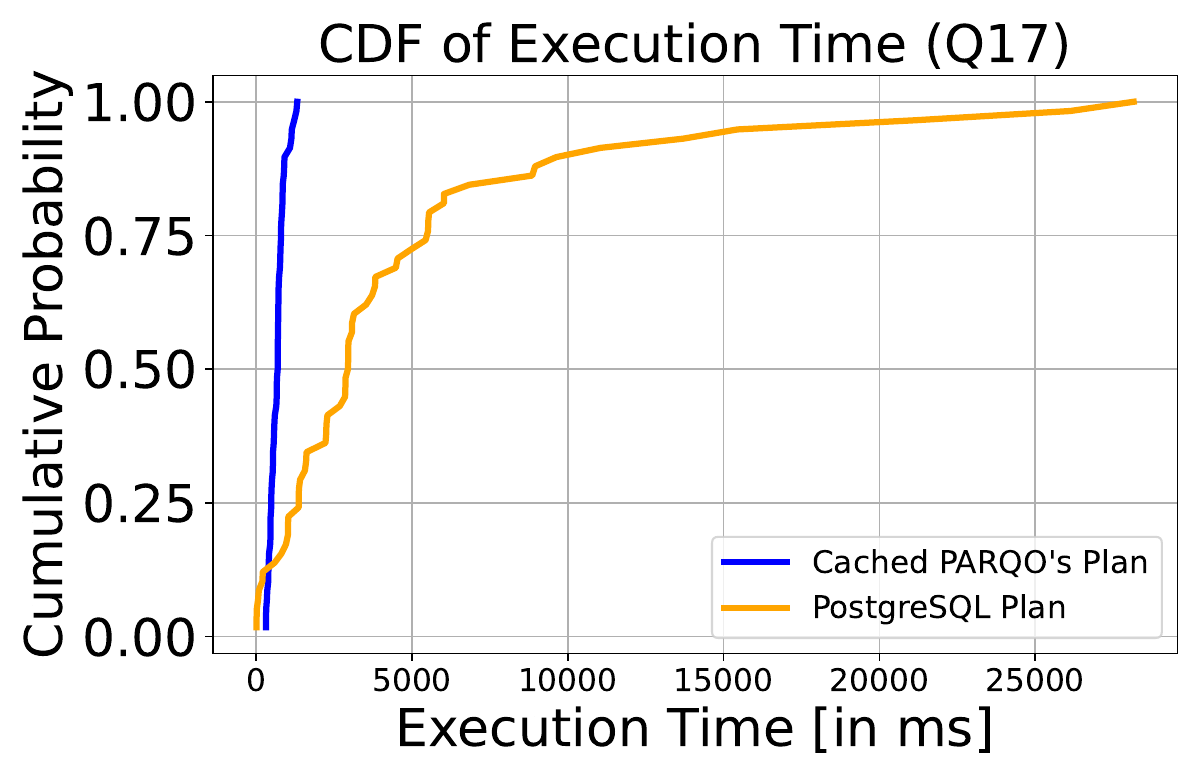}
    \includegraphics[scale=0.15]{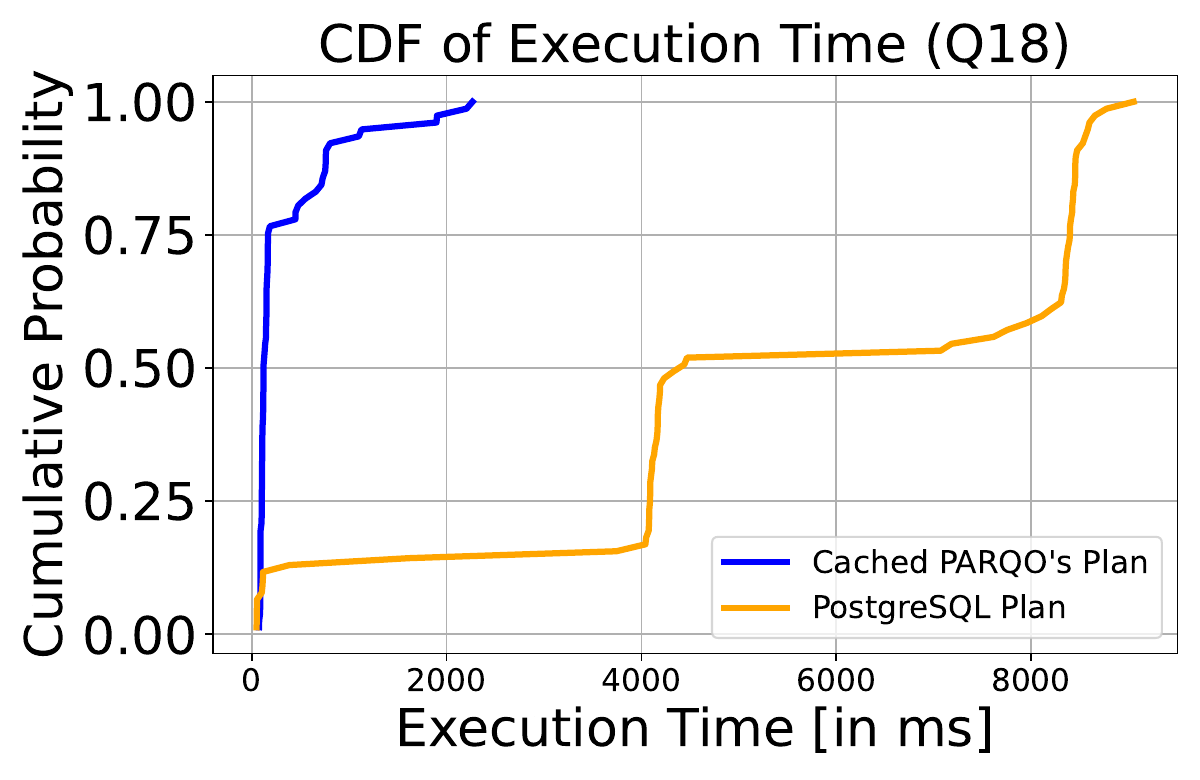}
    \vspace{-3.5mm}
    \caption{\small{\revb{Cumulative density function of execution time in PQO: for Q6, Q4, Q17 and Q18}}}
    \label{fig:latency-cdf}
    \vspace{-3.5mm}
    \end{minipage}

\end{figure*}

We now turn to queries where \OurSys-Sobol underperforms PostgreSQL.
As argued above, a better way of evaluating robust plans is to examine their performance over a range of situations.
Indeed, in later experiments such as PQO, we will see that \OurSys-Sobol plans are robust despite their misfortune on the current database instance.
For example, Q6 is the worst case for \OurSys-Sobol in this experiment, but in the PQO setting we are able to achieve a \common{$2.57\times$} speedup (\Cref{fig:pqo}).
Nonetheless, it is instructive to study why the robust plans underperform in these particular cases.
Delving deeper, we believe the reason lies in the uncertain nature of selectivity estimates.
For instance, Q1 has a sensitive dimension matching $\mathit{it}^\select \join \mathit{mi\_idx}$ with \textit{it.info = `top 250 rank'};
it happens that the estimate was not bad, but our error profiling thought the error would be large.
Similarly, for querylet $\mathit{mk} \join k^\select$ in Q3, 4, and 6, and $\mathit{cn}^\select \join \mathit{mc}^\select$ in Q15,
the actual errors were somewhat unlikely according to our error profiles.
Such discrepancies could arise due to uncertainty, which is inevitable, or due to poor error profiling;
it is hard to tell which on the basis of a few particular cases.
\common{As our later experiments that examine many possibilities in aggregate generally produce good results,
we believe our error profiling is adequate if still imperfect.}

\common{The overall speedup for the entire evaluation workload of JOB is $3.23\times$.
Detailed results for DSB and STATS can be found in~\cite{fullversion}.
As a brief summary here, the overall speedups for all queries in DSB and STATS are $2.01\times$ and $1.36\times$, respectively.
For DSB, \OurSys\ outperforms PostgreSQL in 8 out of 15 queries, with a maximum speedup of $8.8\times$;
for STATS, \OurSys\ outperforms PostgreSQL 10 out of 26 queries, with the highest speedup of $425.7\times$ observed.
The only regression across these benchmarks occurs in S120 ($\downarrow1.87\times$); however, the benefits can still be evident in the PQO experiments.}

\begin{figure*}
    \centering
    \includegraphics[scale=0.14]{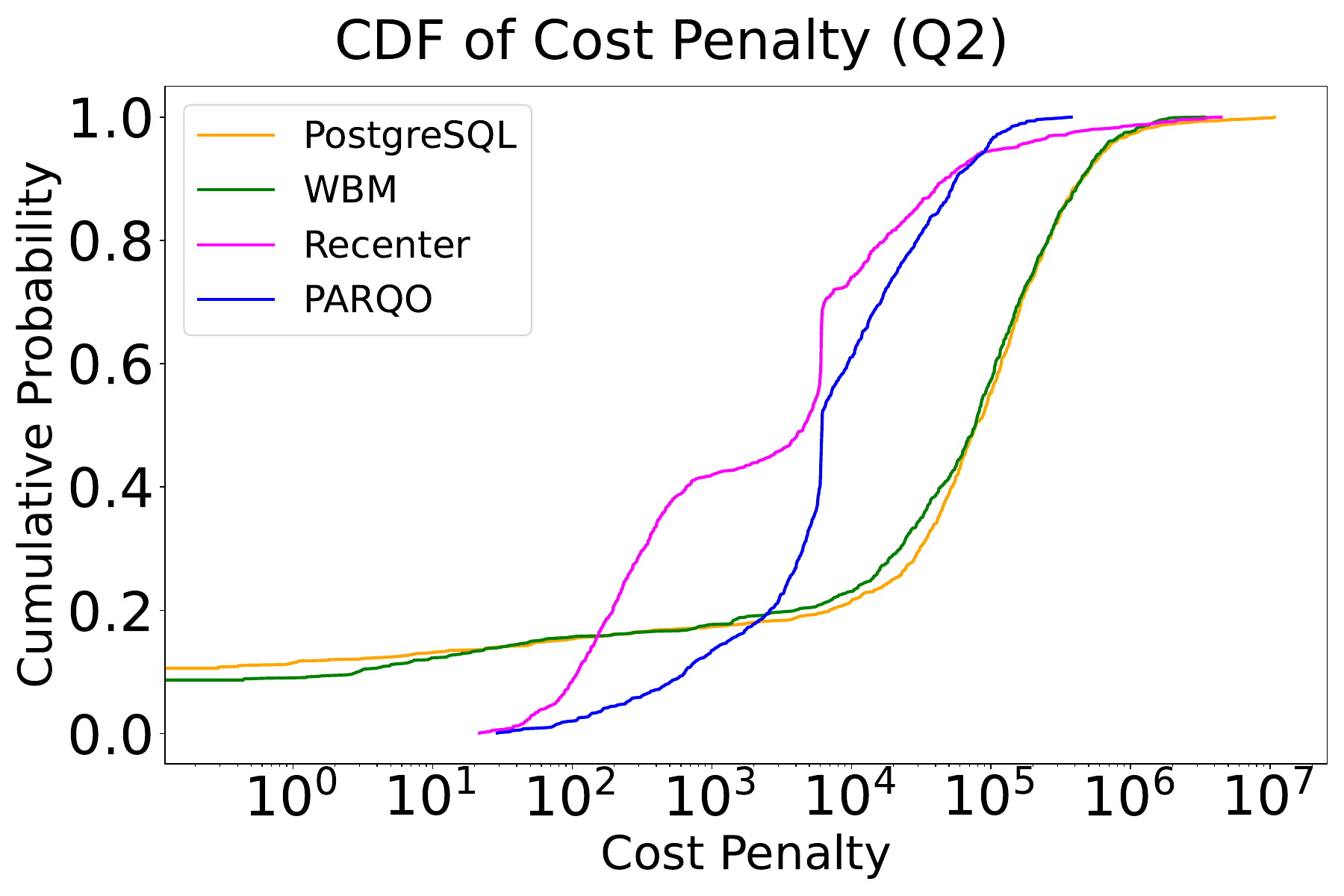}
    \ \ 
    \includegraphics[scale=0.14]{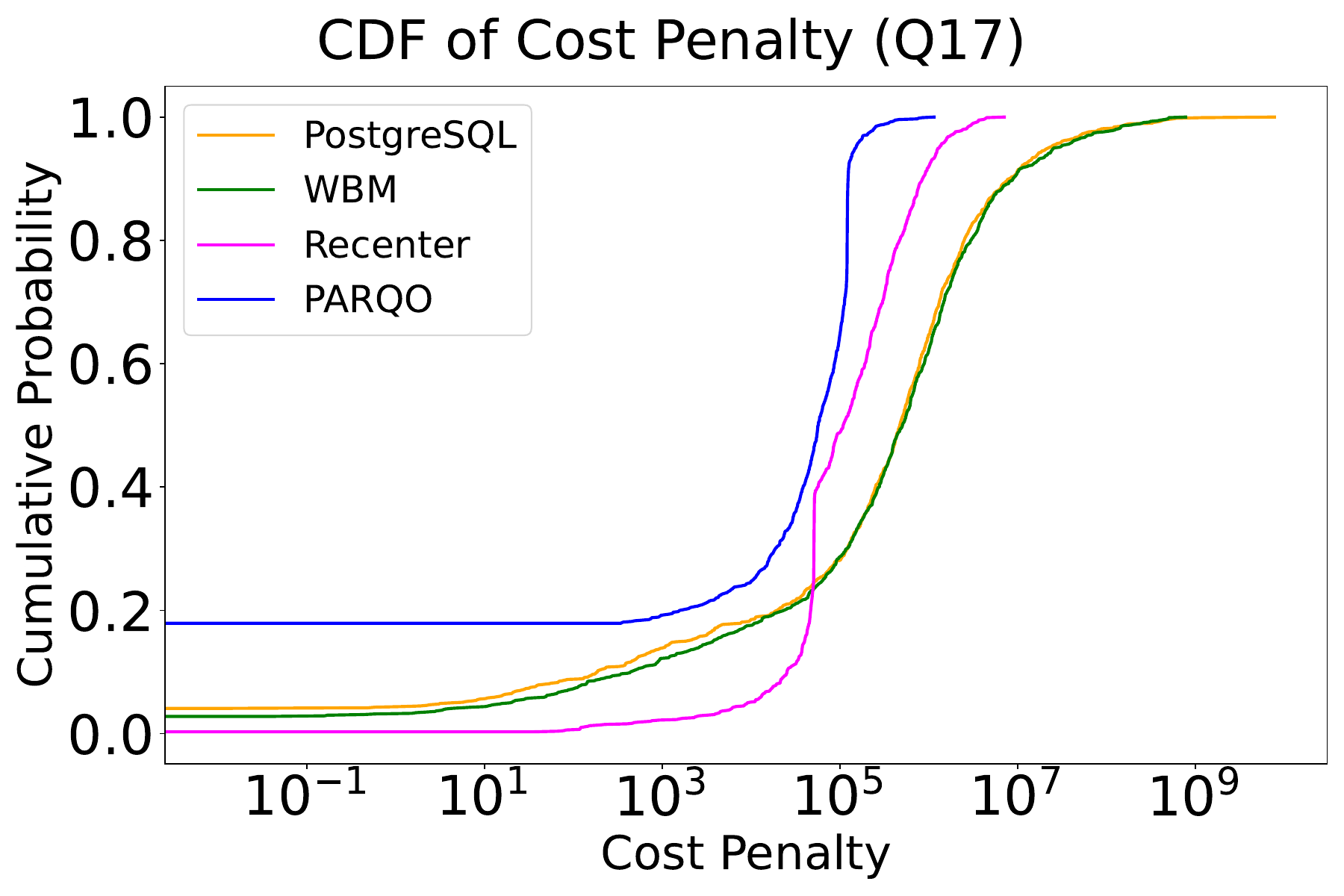}
    \ \
    \includegraphics[scale=0.14]{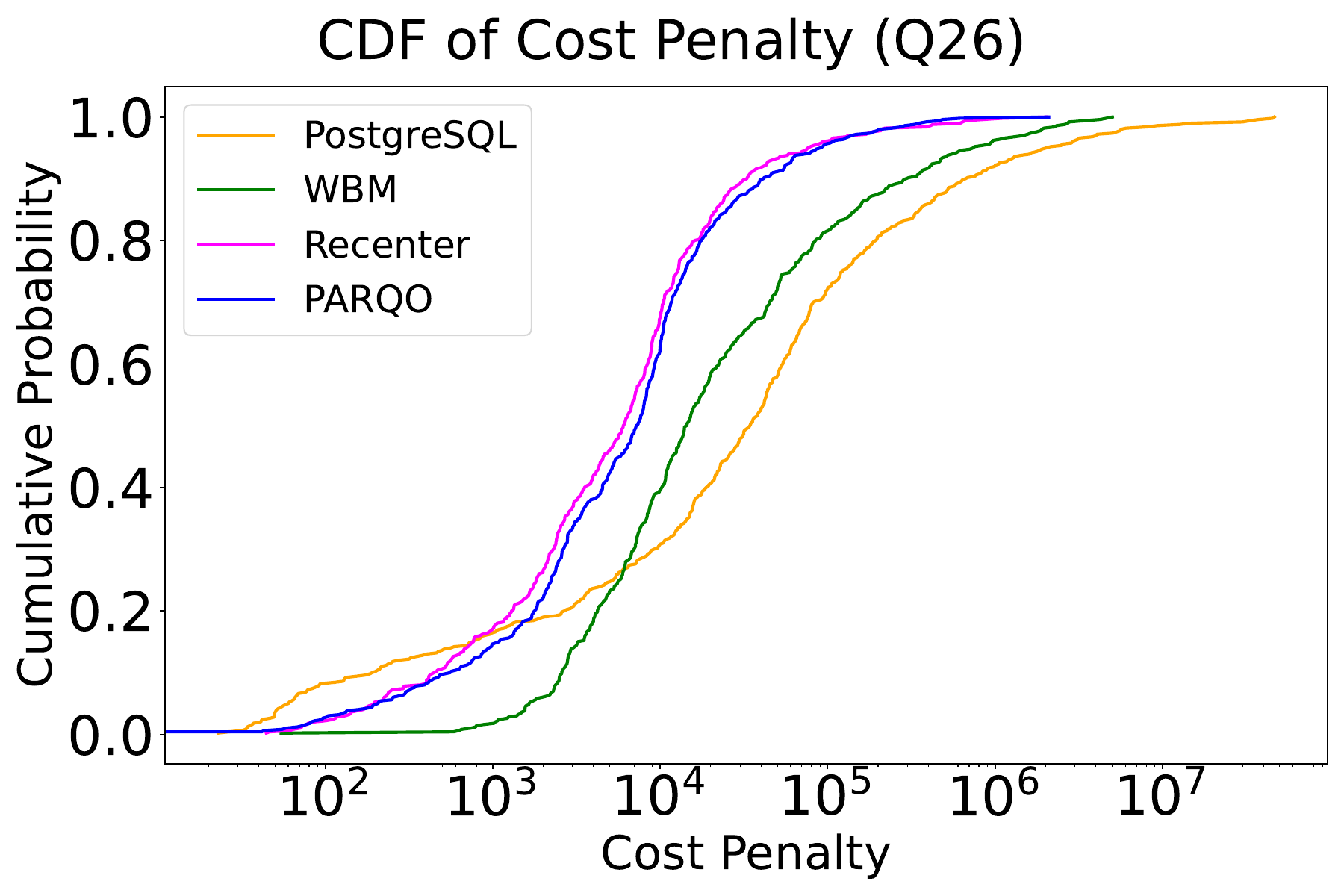}
    \ \
    \includegraphics[scale=0.14]{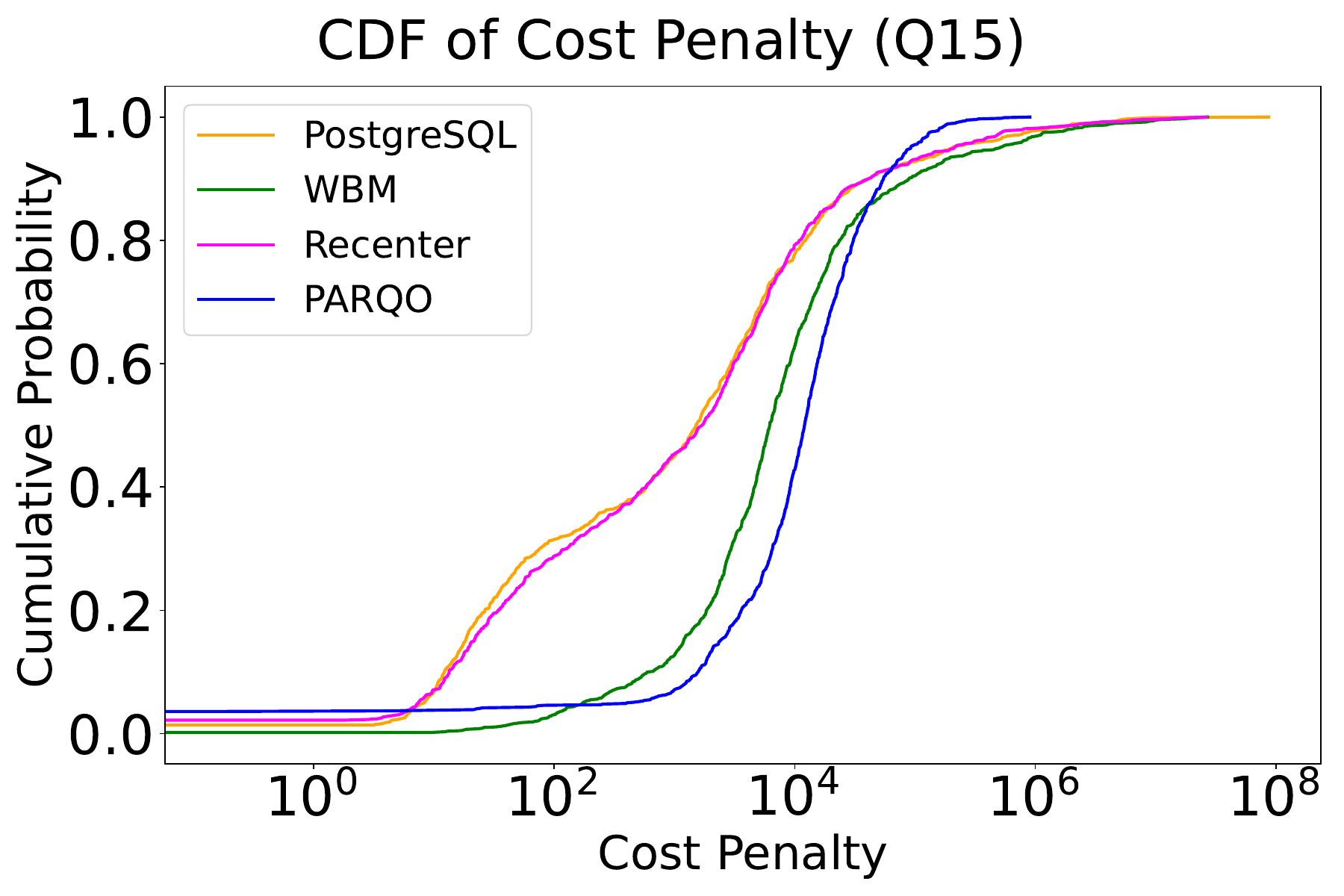}
    \vspace{-4mm}
    \caption{\small{\common{Cumulative density of cost penalty incurred by plans for Q2, 17, 26 and 15 in JOB.}}}
    \label{fig:cdf}
    \vspace{-4mm}
\end{figure*}

\begin{figure*}
    \centering
    \includegraphics[scale=0.105]{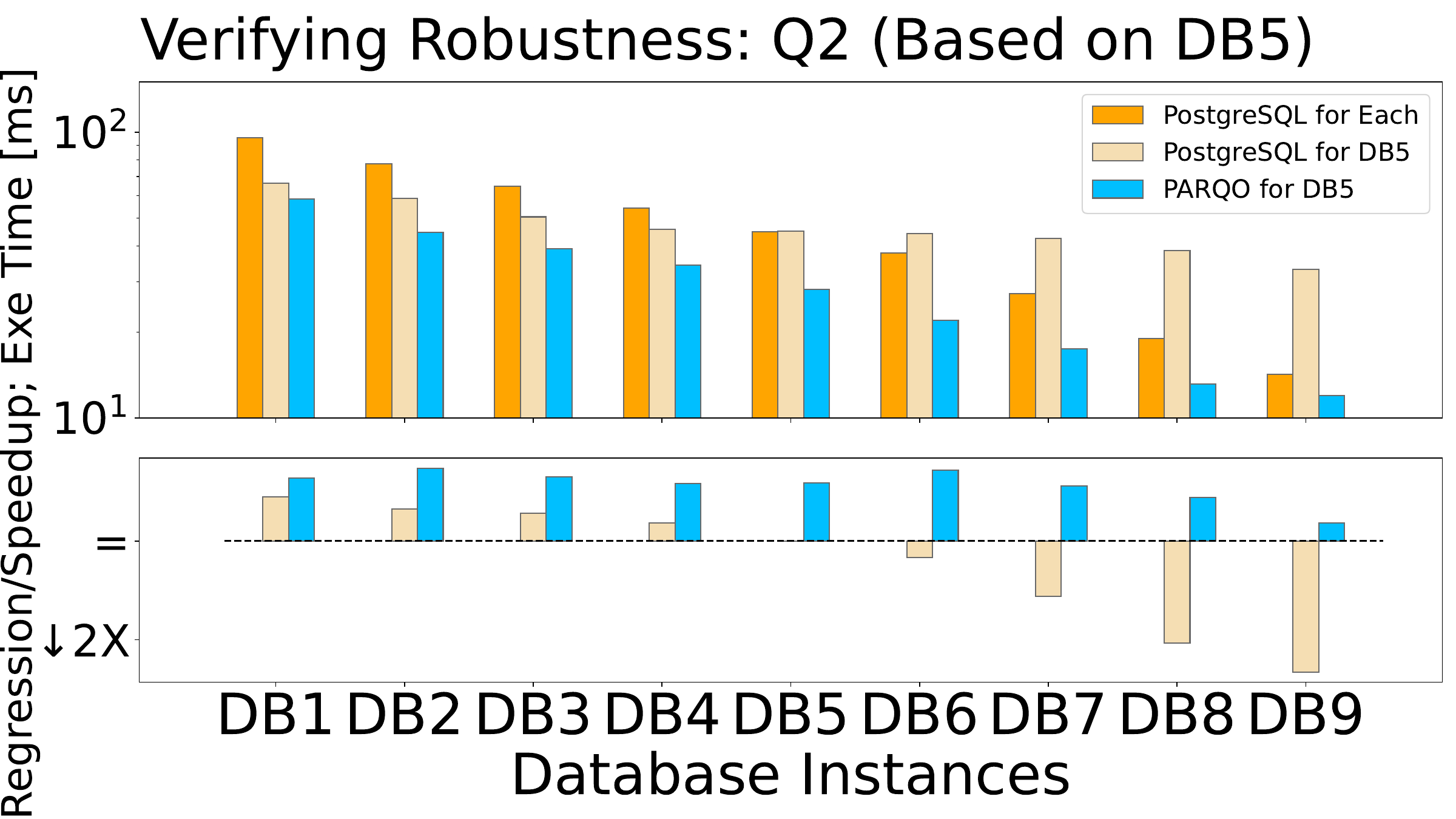}
    \ \ 
    \includegraphics[scale=0.105]{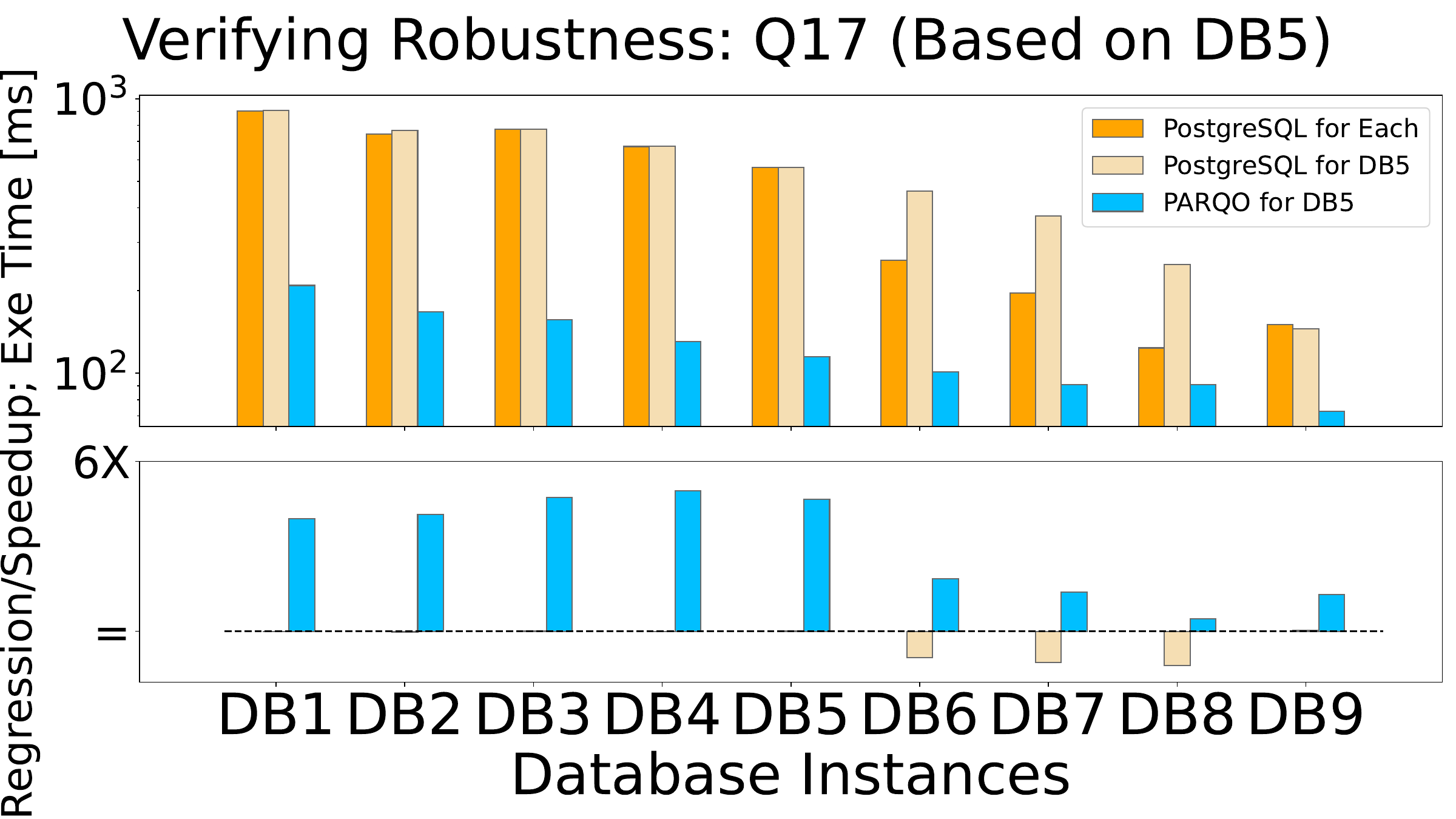}
    \ \
    \includegraphics[scale=0.105]{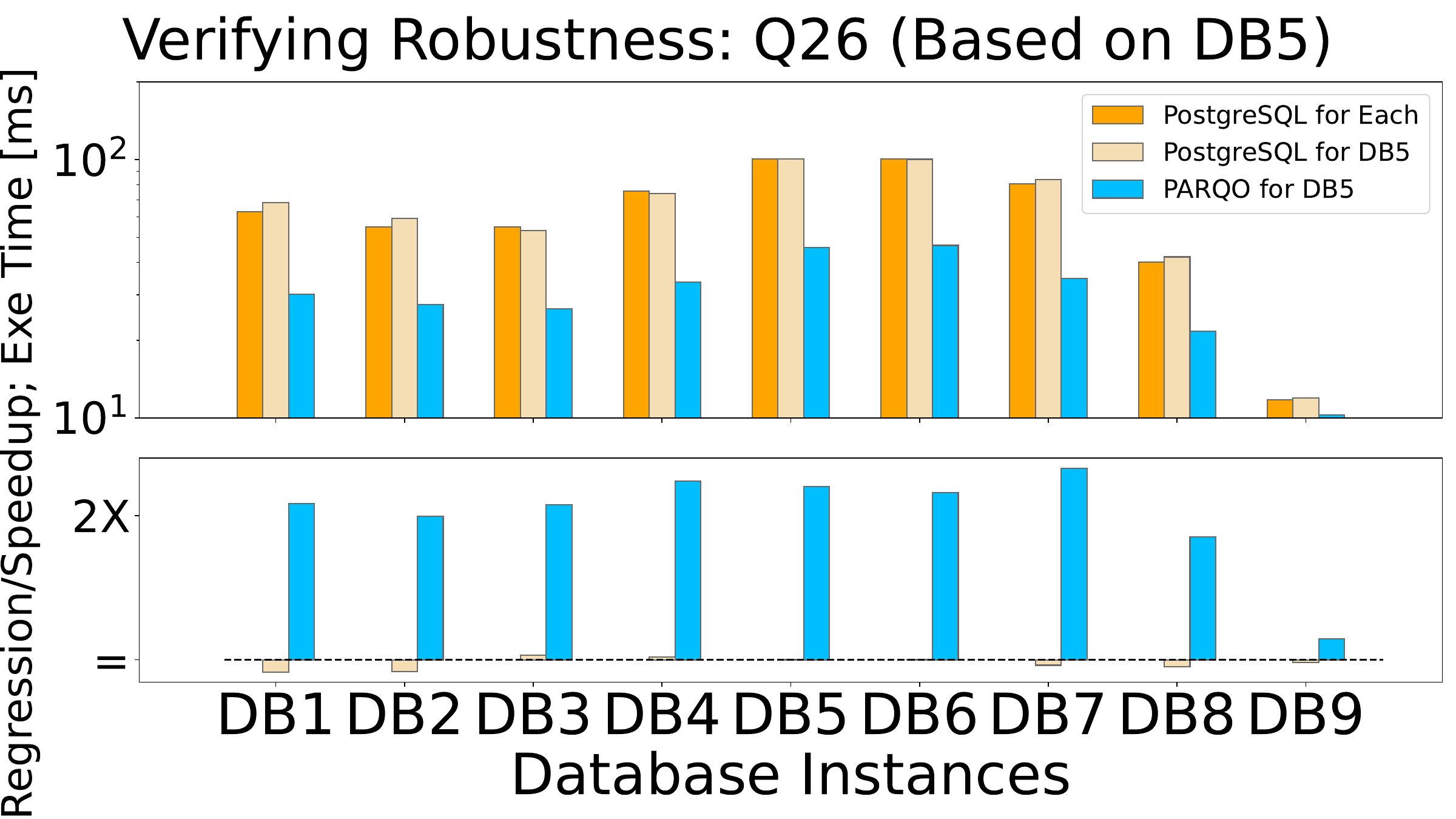}
    \ \
    \includegraphics[scale=0.105]{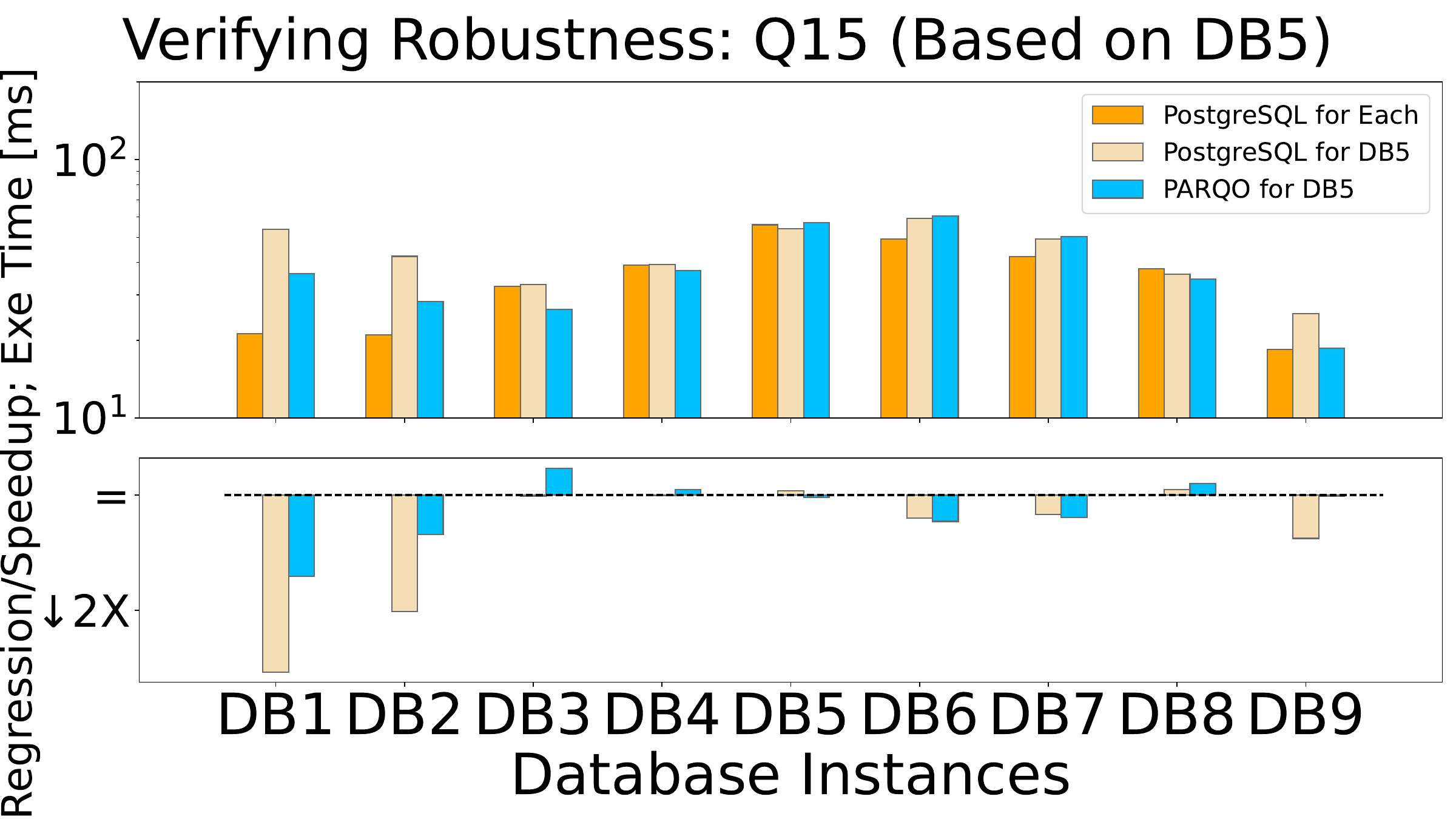}
    \vspace{-4mm}
    \caption{\small{\common{Actual execution times of JOB queries on multiple instances by time-slicing IMDB. DB5 is the base instance.}}}
    \label{fig:verify-slide}
    \vspace{-3mm}
\end{figure*}

\mparagraph{Demonstrating Robustness over Error Distribution}
Next, we demonstrate the robustness of various plans by showing their cost penalties over possible errors in selectivity estimates.
Continuing with the previous setup, for each plan and the initial selectivity estimates $\hat\sels$,
we sample true selectivities $\sels$ according to \Cref{eq:sel-pdf} obtained by our error profiling
(hence, the results here do not validate the quality of error profiling itself),
and cost the plan at $\sels$.
The resulting costs are shown as a cumulative density function.
\Cref{fig:cdf} summarizes the results for JOB.
There are too many queries to show, so we choose four as representative examples: Q2, 17, 26, and 15.
They represent a range of complexities, from simpler 5-table joins to more complex 12-table joins,
and Q15 is intentionally chosen as it showed a regression in our first experiment (\Cref{fig:exe-all}).
\reva{Since \OurSys-Sobol is generally more effective than \OurSys-Morris, we focus on \OurSys-Sobol here.}
As shown in \Cref{fig:cdf}, \OurSys\ plans indeed demonstrate robustness:
they have substantially lower chance of incurring large cost penalties compared with plans selected by alternative methods.
\common{For example, for Q17, PostgreSQL and WBM plans incur $\ge 10^6$ penalty $30\%$ of the time, while the worst-case penalty of \OurSys\ is $10^6$.
Notably, for Q15,
we do see that robustness comes with a price in the low-penalty region:
$30\%$ of the time, PostgreSQL and WBM have penalties $\le 10^2$, while \OurSys\ can reach $10^4$.
However, the protection offered by \OurSys\ shines in the high-penalty region:
PostgreSQL and WBM plans have non-trivial probabilities of incurring catastrophic penalties between $10^6$ and $10^8$,
while \OurSys\ only reaches $10^5$ to $10^6$ in the worst case.}

\mparagraph{Verifying Robustness using Multiple Database Instances}
The above demonstration assumes that estimation errors follow the distribution obtained using our profiling method,
but we also wish to test robustness in less controlled settings encountered in real-world scenarios where additional errors arise as databases evolve.
To simulate such settings, for JOB, which has a static snapshot of the IMDB dataset,
we create multiple database instances by slicing the original dataset into smaller pieces.
We choose one of these as the \emph{base instance}, and apply \OurSys\ (and alternatives) to choose the best plan using information on this instance alone
(e.g., error profiling is done only on this instance).
Then, we execute and time the same plan on the other instances, without knowledge of or regard to selectivities or estimation errors on these instances.
For comparison, we also run the same PostgreSQL plan chosen for the base instance on these instances,
as well as the PostgreSQL plan optimized specifically for each instance (which has the advantage of seeing its statistics).
We use the latter as the reference for speedup/regression factors.

We consider two ways to slice the IMDB dataset used by JOB.
The first is \textbf{\emph{time-slicing}},
where we sort the the title ($t$) rows by \textit{production\_year} and use a sliding window on them to create 9 instances labeled DB1--DB9.
Each instance contains $20\%$ of the title rows along with associated data from other tables,
and two consecutive instances have $10\%$ of the title in common.
The results for the same four representative queries from JOB are shown in \Cref{fig:verify-slide}, with DB5 as the base instance.
For Q2, 17, and 26, we see that the plan chosen by \OurSys\ with the knowledge of DB5 outperforming PostgreSQL plans not only for DB5 but also for all other instances; it even outperforms the instance-optimized PostgreSQL plans, which were obtained with access to better information on their corresponding instances.
For Q15, \OurSys\ is just slightly worse than the base PostgreSQL plan (for DB5) on 3 instances (DB5, DB6, and DB7, which are consecutive in time and may have similar statistics) out of the 9;
however, it is much more robust overall, avoiding the significant performance degradation experienced by the base PostgreSQL plan on DB1, DB2, and DB9.
It is able to outperform the instance-optimized PostgreSQL plans on DB3, DB4, and DB9, despite not having any knowledge about these instances.

The second way to create multiple instances for JOB is \textbf{\emph{category-slicing}},
where we partition the IMDB dataset by item categories (\textit{kind\_type.kind}) such as ``Movie'' and ``TV Series'',
and name each of the 6 result instances by the category.
We intended this partitioning to create more challenging scenarios than time-slicing,
because items in these categories follow very different distributions.
The results, detailed in~\cite{fullversion}, point to similar conclusions as above.

\begin{figure}[t]
    \begin{minipage}[right]{0.37\columnwidth}
    \vspace{3mm}
        \includegraphics[scale=0.102]{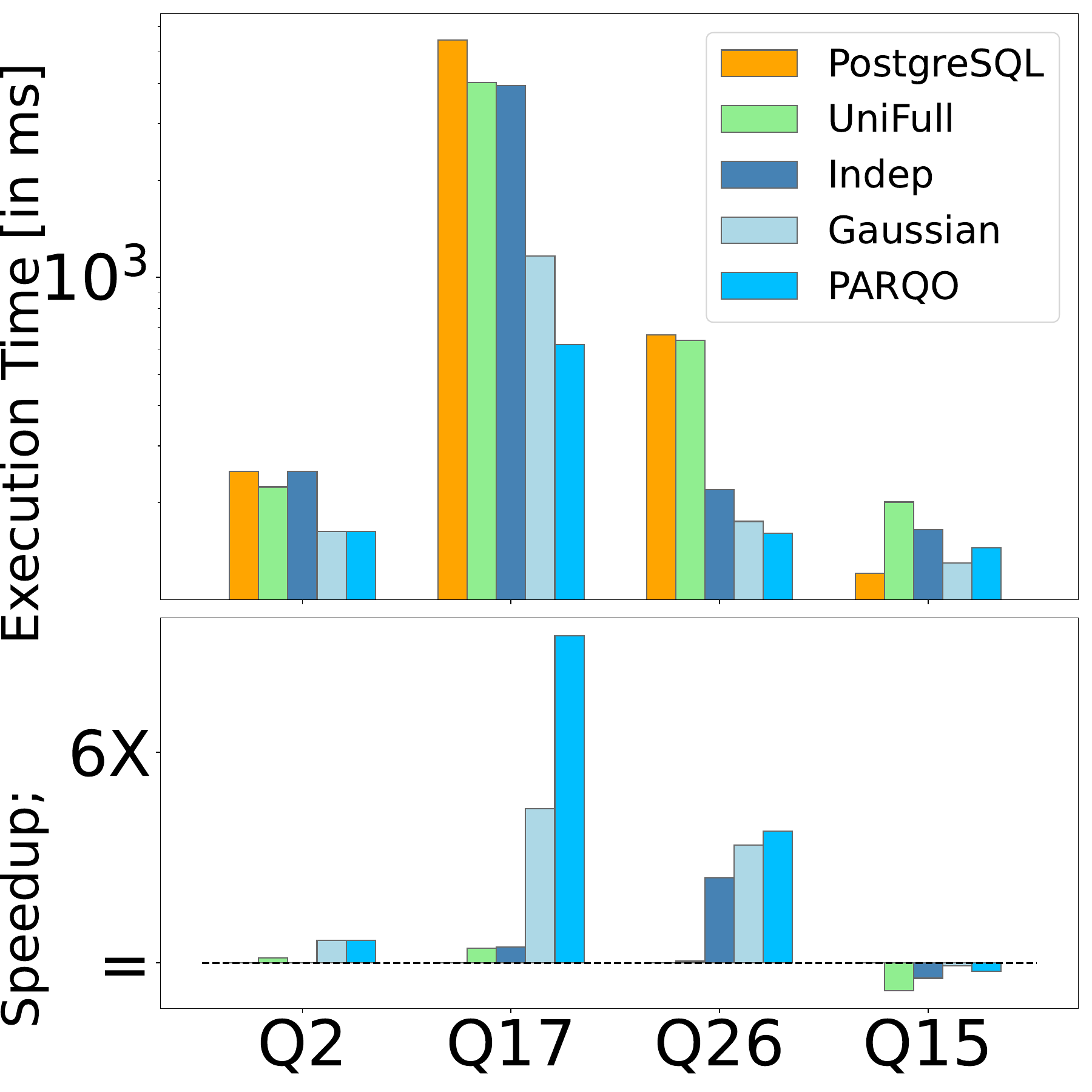}
        \vspace{-5mm}
        \caption{\common{\small{Comparison of error profiling methods.}} }
        \label{fig:exact-and-quality}
    \end{minipage}
    \hfill
    \begin{minipage}[left]{0.6\columnwidth}
    \vspace{-2mm}
    \includegraphics[scale=0.125]{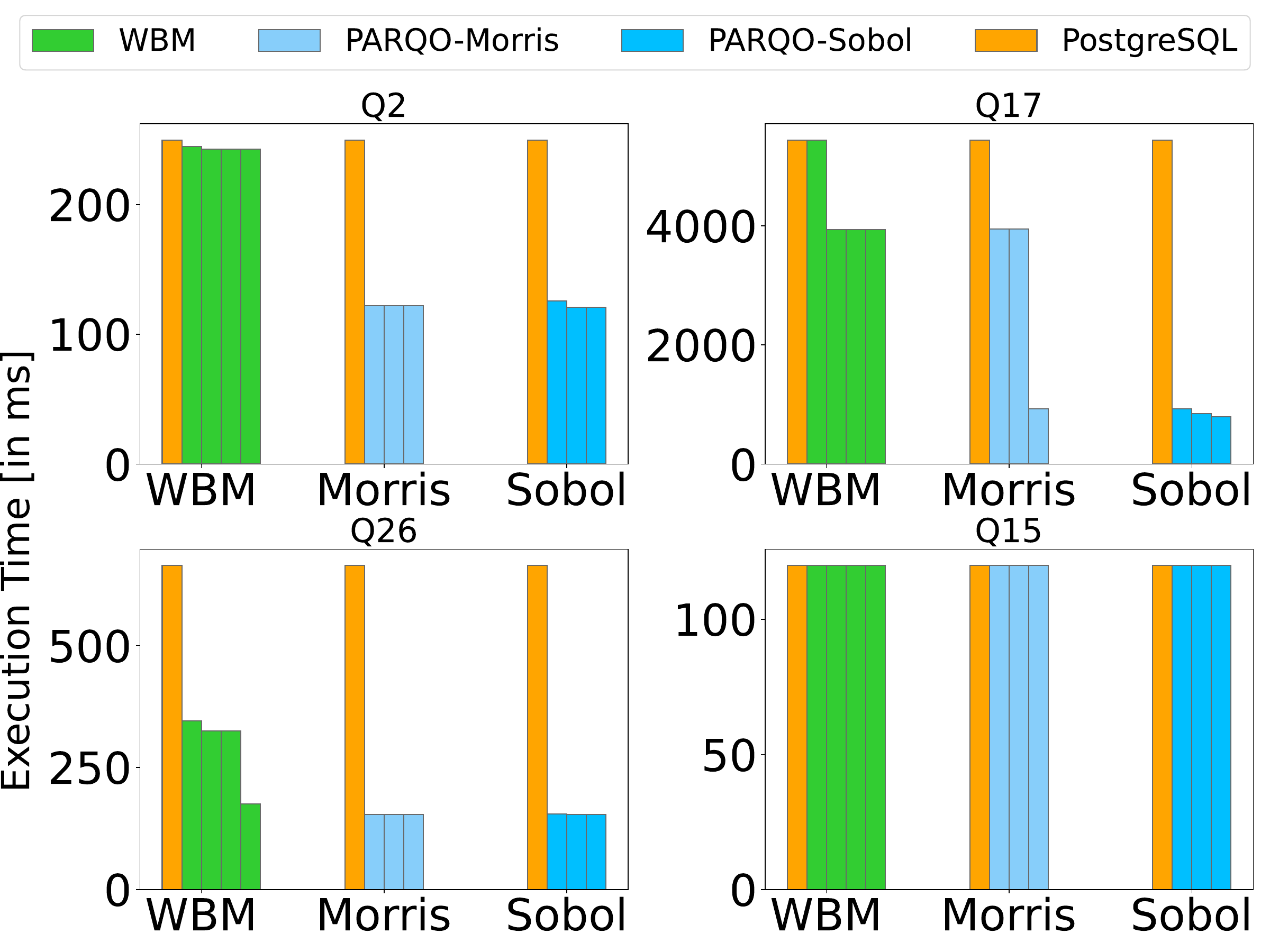}
    \vspace{-7mm}
    \caption{\common{\small{Improvements by correcting estimates for sensitive dimensions.}}}
    \label{fig:inject}
    \end{minipage}
    \vspace{-5mm}
\end{figure}
\vspace{-2mm}
\mparagraph{Impact of Error Profiling Strategies}
We further explore the impact of various approaches to error profiling on \OurSys's performance.
First, the traditional PostgreSQL optimizer can be seen as taking an extremely simple approach of assuming no estimation error.
The second approach, \emph{UniFull}, encodes the assumption in~\cite{abhirama2010stability, chaudhuri2010variance, chu1999least}
that true selectivities are drawn uniformly at random from the entire selectivity space.
\common{The third approach, \emph{Indep}, assumes that join and selection selectivities are estimated independently and their estimation errors are also independent;
hence, we only need to profile errors for each selection and join condition in isolation.}
The fourth approach, \emph{Gaussian}, is identical to \OurSys's method described in \Cref{sec:error},
except that it fits a single Gaussian distribution to each bucket of collected errors instead of using kernel density estimation.
We run \OurSys-Sobol with error models obtained under these strategies to optimize the four representative JOB queries,
and the resulting plans are timed.
The results, shown in Figure \ref{fig:exact-and-quality}, indicate that \emph{UniFull} yields marginal improvement over PostgreSQL,
underscoring the importance of incorporating better knowledge on errors in robust query optimization.
\common{\emph{Indep} does better than \emph{UniFull} but still much worse than \emph{Gaussian} and \OurSys's default method,
highlighting the need to profile dependencies among selectivities as we described in \Cref{sec:error}.
Finally, \emph{Gaussian} further improves upon \emph{Indep} but sometimes underperforms \OurSys's default,
because its single-Gaussian model is crude compared with \OurSys's default.}
\common{For this experiment and all other experiments including those on DSB and STATS,
the memory footprint of \OurSys's error model is always under 15KB.
This low memory usage leaves considerable room for improving model accuracy;
it will be interesting future work to investigate how much additional improvement can be gained with more sophisticated error modeling.}

\mparagraph{\common{Effectiveness of Sensitive Dimensions in Prioritizing Corrections of Estimates}}
\common{To show that \OurSys\ can identify a good set of sensitive dimensions (\Cref{sec:sensitivity}),
we consider the following setup,
motivated by~\cite{chaudhuri2009exact, lee2023analyzing}.}
Given a plan optimized by PostgreSQL with estimates $\hat\sels$, and a list of sensitive dimensions recommended by different methods,
we would acquire the true selectivity values for these dimensions%
\footnote{\label{footnote:get-true-sel}For the purpose of this experiment,
we simply run a \sql{COUNT} subquery for each selectivity of interest,
but in practice one can instruct the database system to refresh statistics relevant to the selectivities or use sampling method to answer the \sql{COUNT} subqueries quickly but approximately.}
and ask PostgreSQL to reoptimize the query based on the accurate selectivities instead of their estimates.
We process the list of sensitive dimensions iteratively and obtain a new plan after correcting one additional dimension at a time;
all plans are executed and timed.
We consider the three most sensitive dimensions found by the Morris and Sobol's Methods, sorted by their sensitivity.
\common{We compare them with WBM's choice of sensitive dimensions, which include all non-key-foreign-key join selectivities;
these dimensions are ordered using $\partial\,\Cost(\plan, \sels) / \partial\sel_i$, based on one of their robustness metrics.
We also note that WBM's behavior in this experiment is not affected by the 120\% threshold.}

\Cref{fig:inject} shows the results on the full IMDB database, with the progression of bars showing how quickly query performance is improved by following each recommendation.
The extra last bar for WBM shows the final plan after processing \emph{all} of WBM's sensitive dimensions.
For Q2, 17, and 26, we see that correcting the top three sensitive dimensions with both Sobol and Morris results in significant speed improvements,
but Sobol ``converges'' quicker.
\common{WBM is only able to match the same improvement for Q26 after correcting all its sensitive dimensions;}
for Q2 and 17, it never reaches the level of Sobol and Morris.
Finally, for Q15, none of the methods improves upon the original PostgreSQL plan, indicating that this plan is already very good for the given database instance.

There is also a trade-off in how expensive these methods are.
The slope-based metric in WBM only require $d$-1 calls to \Opt, as it is local and OAT.
Morris and Sobol perform better but require far more \Opt\ calls.
As an example, for Q17, Morris requires 520 calls ($K=40$) to its solution, while Sobol requires 1{,}664 calls ($K=64$).
In fact, the cost of finding sensitive dimensions dominates that of robust query optimization ---
once the sensitive dimensions are identified, finding the robust plans only requires additional $S=100$ \Opt\ calls
(we also experimented with $S=1{,}000$ but did not find obvious improvement in overall performance).
\common{The total overhead is considerable, averaging at several minutes per query, which renders the approach applicable only to very slow queries.
Luckily, the complexity of robust query optimization depends only on query complexity and not on data complexity,
so it is more appealing to massive databases.
For faster queries, instead of sacrificing solution quality and principality,
we argue for combining robust query optimization with parametric query optimization,
such that the overhead of optimization is amortized over many queries sharing the template.}
Next, we present results from the PQO experiments, along with a more detailed analysis of overhead.

    

\mparagraph{Parametric Query Optimization}
The PQO experiment setup for JOB requires a bigger evaluation workload of queries beyond the 113 included in the benchmark.
Here, we use all 113 queries as our profile workload and collect all literals therein by their attribute domains.
The multiset of literals from the same domain defines the distribution to be used when generating new queries requiring literals from this domain.
For each of the 33 templates, we generate 1000 random query instances for the evaluation workload,
where each literal is replaced with one randomly drawn from the same domain.
We treat the 33 JOB queries labeled (a) as anchors, perform robust query optimization on each,
and populate the PQO plan cache with 100 samples and up to 3 robust plan candidates obtained when optimizing the anchor (\Cref{sec:robust:pqo}).
While it is certainly possible to use more than one anchor per template or to cache more per anchor,
we have found this modest level is already sufficient to achieve satisfactory performance.

\common{Overall, the total time for PostgreSQL to execute the entire evaluation workload of 33{,}000 queries is 6.59 hours.
\OurSys's PQO setup reduces this time to 4.34 hours (not yet including the upfront overhead of populating the cache, which we discuss later).}
\Cref{fig:pqo} summarizes the results by query template.
We use the term \emph{reuse fraction} to represent the proportion of queries that trigger reuse
(by passing the KL-divergence test with respect to the anchor associated with its template).
The average reuse fraction over all templates is \revb{37\%}.
For each template, \Cref{fig:pqo} compares the average execution time over queries that reused a cached \OurSys\ plan
against plans that PostgreSQL chooses.
We omit templates with a reuse fraction below 5\% (\revb{Q10 and 33}) because the numbers are too low to draw reliable conclusions.%
\footnote{For these and other templates with relatively low reuse fractions,
it seems that their anchors' selectivities are quite different from most of the queries from the evaluation workload.
A smarter way of picking anchors that adjusts to the query workload should be a helpful future work direction.}
Among the remaining 31 templates, \OurSys\ plans achieve a speedup in 28 of them.
Notably, template Q18 has a \revb{$17.8\times$} speedup.
For the 3 templates with no speedup, Q4, 5, and 14, the worst regression is only $\downarrow1.1\times$.
Recall that besides Q4, in the earlier experiment in \Cref{fig:exe-bad}, \OurSys\ also underperformed PostgreSQL on Q1, 3, 6, and 15;
however, here with PQO, we see that queries with templates Q1, 3, 6, and 15 have an overall speedup.

Additionally, \Cref{fig:latency-cdf} shows the distribution of query execution times for queries in four representative templates.
The four are chosen for different reasons.
Q6 was the ``worst'' query for \OurSys\ in the earlier experiment in \Cref{fig:exe-bad}.
Here, we see that while PostgreSQL is slightly faster than \OurSys\ for \revb{$20\%$} of the queries with running time below \revb{250ms},
PostgreSQL causes \revb{$15\%$} of the queries to experience substantially longer running times than \OurSys;
overall \OurSys\ in fact provides a significant speedup (\Cref{fig:pqo}).
Next, Q4 is the ``worst'' query template for \OurSys\ in \Cref{fig:pqo}.
Even in this case, its execution time distribution provides protection against some long-running times incurred by PostgreSQL.
Finally, Q17 is also one of the representative queries chosen in earlier experiments, and Q18 is the ``best'' query template for \OurSys\ in \Cref{fig:pqo}.
As can be seen from \Cref{fig:latency-cdf}, PostgreSQL oftentimes makes disastrous choices for queries with these two templates,
yet \OurSys's robust plans help avoid these situations.

\common{In closing, we present a detailed analysis of the various overheads incurred by \OurSys\ in robust PQO compared with traditional query optimization.
First, for JOB, \OurSys\ incurs a one-time, upfront cost of 2.13 hours to populate its PQO cache for all 33 query templates, which averages at about 4 minutes per template.
While a considerable amount of overhead, it depends on the complexity of the query rather than the size of data,
and if the query template is used often, the initial investment pays off quickly.
Recall that PostgreSQL runs the entire evaluation workload in 6.59 hours while \OurSys\ runs it in 4.34 hours;
the saving of 2.25 hours is already more than the initial overhead of 2.13 \shep{hours}.
A simple calculation reveals that it takes on average about 934 JOB query instances per template to break even the upfront optimization cost.
This target is not difficult to reach in production settings where there is a database application with a limited set of query templates and many users,
or when queries are more expensive than the benchmark setting we experimented with.}
Delving deeper, \OurSys\ saves time not only by having better plans, but also by reducing runtime optimization overhead.
\OurSys's runtime overhead depends on the number of cached plans and samples.
Under our experimental setting, over all query instances that triggered reuse,
\OurSys\ has an average optimization overhead of \revb{5.58ms} per query,
which is much lower than PostgreSQL's optimization overhead of \revb{14.9ms}.
Over all queries instances, \OurSys\ has an average optimization overhead of \revb{55.6ms}
(including the time spent on KL-divergence testing and the time to fall back to PostgreSQL when the test fails),
which is still better than PostgreSQL's average optimization time of \revb{58.8ms}.
There are ample opportunities for future work on selectively picking anchors for PGO to maximize reuse and avoid those with low reuse.

\common{Finally, we briefly summarize the results of PQO experiments on DSB and STATS; futher details are available in~\cite{fullversion}.
The average reuse fraction over all templates is 93\% for DSB and 43\% for STATS. 
Overall, \OurSys~ achieves improvements of 0.68 hours and 11.47 hours in executing the entire evaluation workload for DSB and STATS respectively. 
Additionally, it reduces optimization overhead by 0.09 hours for DSB and 0.01 hours for STATS.}

\begin{table}[t]
    \centering
    \footnotesize 
    \setlength{\tabcolsep}{5pt} 
    \renewcommand{\arraystretch}{0.9} 
    \begin{tabular}{lccc}
        \toprule
        & \textbf{JOB} & \textbf{DSB} & \textbf{STATS} \\
        \midrule
        \textbf{\# of samples $S$ } & 100 & 100 & 100 \\
        \textbf{Physical size of $f(\sels | \hat\sels)$ } & 13.8 KB & 13.66 KB & 5.84 KB \\
        \textbf{\# of relevant dimensions $d$} & 6-34 & 8-25 & 6-20 \\
        \textbf{\# of sensitive dimensions} & 2-6 & 2-4 & 1-4 \\
        \textbf{\# of seeds $K$ of Morris} & 20-160 & 10-80 & 20-80 \\
        \textbf{\# of seeds $K$ of Sobol} & 8-128 & 8-64 & 8-64 \\
        \textbf{Avg \# of unique plans $\grave S$ } & 18 & 14 & 10 \\
        \textbf{Up-front overhead of \OurSys~} & 2.13 h & 0.99 h & 0.74 h \\
        \textbf{Overall speedup per query} & $3.23\times$ & $2.01\times$ & $1.36\times$ \\
        \textbf{Workload size in PQO} & 33,000 & 15,000 & 22,000 \\
        \textbf{Average reuse fraction} & 37\% & 93\% & 43\% \\
        \textbf{Execution time saved by PQO} & 2.25 h & 0.68 h & 11.47 h \\
        \textbf{Optimization time saved by PQO} & 0.03 h & 0.09 h & 0.01 h \\
        \bottomrule
    \end{tabular}
    \caption{\small \common{Summary of \OurSys~ on three benchmarks.}}
    \label{tab:performance_metrics}
    \vspace{-10mm}
\end{table}

\vspace{-1mm}
\section{Related Work}
\label{sec:related}

\textit{Robust Query Optimization} 
How to improve the ability of error resistant and avoid sub-optimal risks has been widely discussed \cite{borovica2017robust, graefe2012robust, haritsa2019robust}. RQO can be regarded as part of robust query processing and is classified by the number of plan provided by a recent survey \cite{yin2015robust}. 
For single-plan based approach, LEC \cite{chu1999least} was among the first to utilize probability density for estimating selectivity, aiming to identify plans with the lowest expected cost. However, LEC restricts the search space of plans and lacks a clear methodology for constructing probability measures.
Similarly, \cite{abhirama2010stability, chaudhuri2010variance} pick a plan that has low variance and minimum average cost over extremes for the entire parameter space. 
Since the error can not be captured in an arbitrary large selectivity space, their effectiveness is limited. 
\reva{In contrast, RCE \cite{babcock2005towards} tries to quantify the uncertainty, but requires random samples from real data at runtime to infer the distribution of actual selectivity.
}
Rio \cite{babu2005proactive} and its extension \cite{ergenc2007robust, alyoubi2016database} leverage bounding boxes or intervals to quantify selectivities, and collect the plan as a candidate if the cost is close to optimal over the bounding box. When executing the query, these candidate plans can be utilized as re-optimization alternatives. 
\revc{The idea of adaptive processing, i.e. collecting running time observations and then switching the current plan to another is also leveraged in \cite{dutt2014plan, dutt2014quest, trummer2021skinnerdb, wolf2018calculation}.} 
\revc{\cite{wolf2018calculation} identifies cardinality "ranges" where the original plan remains optimal. When the "ranges" are broken during execution, re-optimization will be triggered.  This line of work generally requires runtime adaptation and is complementary to our approach, which focuses on compile-time optimization of standard execution plans.}
These interval-based approaches need to assume that predicate selectivity is known with in a narrow intervals, which is often violated in practical situation \cite{job, small}.
Besides, research on plan diagrams \cite{harish2007production, jayant2008identifying} aims to identify a fine-grained set of plan candidates for a query template across the selectivity space, each candidate can be regarded as a robust plan for certain area. Subsequent works \cite{purandaredimensionality, dey2008efficiently} present methods to reduce the complexity of the diagram, but they still necessitate time-consuming offline training through repeated invocations of $\Opt$. Additionally, their plan selection is still reliant on $\hat \sels$, which may lead to sub-optimal outcomes. 
Risk Score \cite{phdthesis2016} employs the coefficient of variation to measure the robustness of execution plans. However, this metric requires real execution times under various actual selectivity values. 
MSO \cite{dutt2014plan, purandaredimensionality, jayant2008identifying, karthik2018concave} is widely used in robust query processing that quantifies the worst-case sub-optimality ratio across the entire selectivity space. It relies on the availability of the real optimal plan, which is typically only known to the optimizer after the query execution has begun. 
\reva{\cite{marcus2021bao, li2023dbet} learn from real executions to make the \textit{optimizer} more robust by improving the mapping between ``plan'' to ``execution time'', and
DbET \cite{li2023dbet} shares a similar idea that predicts the latency of a plan as a distribution. 
Our approach differs in that we aim to identify sensitive cardinality uncertainties and select robust plans in a more explainable manner.
}
\cite{lee2023analyzing} analyzes the tolerance of a plan to cardinality errors posteriorly, requiring true cardinality for all sub-queries and extensive real execution. \shep{In paper, we present a principled approach to access sensitivity instead.}
WBM \cite{2018rqo} presents three alternative metrics (based on the slope or integral of the cost function) that only based on estimation to measure the robustness. 
\common{The robustness metrics in \OurSys~ follow this direction that are only based on estimation without executing the query or subquery and independently evaluate each plan. }

\textit{Parametric Query Optimization}
PQO has been a subject of study for three decades \cite{hulgeri2002parametric, ioannidis1997parametric}. 
The primary focus is to minimize the optimizer's invocation by utilizing cached plans while ensuring the plan's cost remains acceptable. 
According to \cite{dutt2017leveraging}, current PQO methods can be classified by the plan identification phase, which includes online and offline-based methods. 
\revc{Online-based methods are widely used in commercial DBMS ~\cite{microsoftQueryStore}.} \cite{alucc2012parametric, ghosh2002plan} build a density map by clustering executed queries to select stored historical plans for new query. 
\revc{Idea of storing the optimality ranges for plans \cite{wolf2018calculation} can also be applied.}
A recent study \cite{dutt2017leveraging} introduces "re-cost" to efficiently calculate $\Cost(\plan, \sels)$, thereby reducing overhead. "re-cost" is demonstrated effective~\cite{dey2008efficiently} and also employed in \OurSys.
For offline-based methods, the objective is to identify a set of plans work for the entire selectivity space \cite{hulgeri2002parametric, ioannidis1997parametric}. Plan diagram \cite{harish2007production, picaso} is applicable in this setting.
A novel framework \cite{vaidya2021leveraging, doshi2023kepler} 
uses actual execution times for all candidate plans to train a model for each template and predict the best plan for new queries. Candidate plans are searched from executed queries \cite{vaidya2021leveraging}, or generated by randomly perturbing different dimensions \cite{doshi2023kepler}, which is similar to our candidate plans searching. However, \OurSys~ focuses on sensitive dimensions and samples from $f(\sels | \hat\sels)$ directly. \cite{vaidya2021leveraging, doshi2023kepler} demonstrate that learning from real executions can accelerate PQO, offering a promising avenue for future research.
\common{As shown in \cref{sec:expr}, \OurSys~ can serve as an offline plan caching and online plan selection method, and it can readily be extended to cover the entire selectivity space using techniques such as clustering \cite{alucc2012parametric, ghosh2002plan} or plan diagrams \cite{picaso}. Our experiments demonstrate that robust plans are effective in PQO settings. Even without necessitating real execution times of query instances like \cite{doshi2023kepler}, \OurSys~ enhances the overall performances.}

\vspace{-3mm}
\section{Conclusion and Future Work}
\label{sec:conclude}


Besides various future work directions already mentioned earlier
(such as better error profiling and visualization/interfaces aided by sensitive dimensions),
we outline several more below.
First, we still do not have a theoretical guarantee on \OurSys's solution optimality with respect to robustness.
We feel that principled sensitivity analysis proposed by \OurSys\ is a promising approach to the problem
from the angle of reducing dimensionality, but more work is still needed in this direction.
Another angle that needs to be further investigated in order to achieve any optimality guarantee
is the identification of robust plan candidates.
Our current approach intuitively looks for candidates among optimal plans at different points in the selectivity space,
but what if the most robust plan is not optimal (or even among the top optimal) for any single point?
New methods are needed for surfacing such elusive candidates and/or determining that they are unlikely to exist.
To make progress, we may need to reconsider the limited ways of interacting with existing query optimizers
(which was done by \OurSys\ for practicality),
and instead seek tighter integration with the optimizer core.
Finally, while having an error distribution enables stochastic optimization,
what if the error distribution changes?
It can be argued that whenever we notice a significant change in error distribution,
the first course of action should be to refresh statistics and retrain models.
If that first line of defense is able to restore the error distribution back to acceptable levels,
it will help make changes to error distribution smaller or less frequent.
Some of the techniques we already employ in \OurSys\ (e.g., KL-divergence tests, caching, and importance sampling)
can help detect changes and lower the cost of adaptation,
but a comprehensive solution for handling such changes still needs to be developed and evaluated.

\bibliographystyle{ACM-Reference-Format}
\bibliography{sample}


\clearpage

\appendix
\section{Alternative Notions of Robustness}
\label{app:alt-robust}

\shep{%
As motivated in \Cref{sec:introduction}, we do not wish to dictate a single notion of robustness because it depends on the application scenario.
A system that specializes in one notion but ignores others will not be able to gain wide adoption.
Hence, the philosophy of \OurSys\ is to provide a flexible framework that supports different user-defined notions of robustness.
To illustrate the generality of our framework, we briefly discuss several alternative notions of robustness below, besides the one defined using~\Cref{eq:penalty}.
\begin{itemize}[leftmargin=*]
\item \emph{Probability that the cost exceeds a given tolerance factor of the optimal.}
    Lower probabilities means a more robust plan.
    \begin{equation}\label{eq:probability}
        \Penalty(\plan, \sels) = \mathbf{1} (\Cost(\plan, \sels) > (1 + \tau) \cdot \optCost(\sels)).
    \end{equation}
    Here, $\tau$ specifies the tolerance factor, and note that the expectation of the above yields the probability.
\item \emph{Variance in the extra cost incurred beyond the optimal.}
    Lower variance means a more robust plan.
    \begin{equation}\label{eq:sde}
    \begin{aligned}
        \Penalty(\plan, \sels) &= \left(\Cost(\plan, \sels) - \optCost(\sels) - \mu\right)^2\\
        \text{where}\; \mu &= \Expectation[\Cost(\plan, \sels) - \optCost(\sels) \mid \hat\sels].
    \end{aligned}
    \end{equation}
    The expectation of the above yields the variance in $\Cost(\plan, \sels) - \optCost(\sels)$.
    Since standard deviation is the square root of variance, minimizing variance is equivalent to minimizing standard deviation.
    Note that although the term $\mu$ in~\Cref{eq:sde} involves an expectation itself,
    overall, we can rewrite the expected penalty equivalently as:
    \begin{equation*}
        \Expectation[(\Cost(\plan, \sels) - \optCost(\sels))^2 \mid \hat\sels] - 
        \left(\Expectation[\Cost(\plan, \sels) - \optCost(\sels) \mid \hat\sels]\right)^2,
    \end{equation*}
    where both expectations can be computed efficiently using the same sampling procedures in~\Cref{sec:sensitivity,sec:robust}.
\item \emph{Absolute or relative difference from the optimal cost.}
    As discussed earlier in~\Cref{footnote:other-penalties},
    these are readily expressible in our framework,
    but are only ``pseudo-dependent'' on the optimal cost and hence easier to handle than other notions of robustness.
\end{itemize}

Many previously proposed robustness notions (see~\cite{yin2015robust} for a survey)
only consider the performance of a given plan but not its performance relative to the optimal plan.
For example, \cite{chu1999least} defines the most robust plan as the one with the lowest expected cost,
and \cite{chaudhuri2010variance} considers both expected cost and cost variance;
both are independent of the optimal costs.
Among the three robustness metrics in~\cite{2018rqo},
\emph{Cardinality-Slope} and \emph{Selectivity-Slope} are defined by
summing the partial derivatives of the cost of the given plan with respect to each of its sensitive dimensions at $\hat\sels$;
since they ignore uncertainty in $\hat\sels$, we can encode them in our framework by concentrating all density of $f(\sels | \hat\sels)$ at $\hat\sels$.
The third, \emph{Cardinality-Integral}, is defined by
summing the integrals of the cost of the given plan over each of its sensitive dimensions;
we can encode this integral as the expected penalty by letting $f(\sels | \hat\sels)$
randomly select one sensitive dimension and then a value for this dimension uniformly at random.
Again, all three metrics in~\cite{2018rqo} are independent of the optimal plans.
We experimentally validate the advantage of our approach over~\cite{2018rqo} in \Cref{sec:expr}.

\mparagraph{Extending Robustness Beyond Expected Penalty}

All notions of robustness described above fit in the optimization objective of \Cref{eq:expected-penalty}%
---i.e., minimizing expected penalty---%
simply by plugging in different definitions of $\Penalty$ and $f(\sels | \hat\sels)$.
We note that we can also extend \OurSys\ to support other notions of robustness that are not traditional stochastic optimization objectives.
To illustrate, consider the following example.
As mentioned in \Cref{sec:introduction},
many previous robust optimization approaches do not leverage knowledge of the error distribution.
Instead, they look for a plan with the lowest worst-case cost over all possible true selectivities,
which often ends up being an over-conservative choice.
To achieve a similar goal in a less conservative manner,
we can replace \Cref{eq:expected-penalty} with a minimization objective that leverages $f(\sels | \hat\sels)$:
\begin{equation}\label{eq:worst-in-hdr}
\max_{\sels \in \mathrm{HDR}_{0.9}(f(\sels | \hat \sels))} \Penalty(\plan, \sels).
\end{equation}
Here, $\mathrm{HDR}_{0.9}$ denotes 90\% highest-density region of the distribution;
the objective finds the plan that minimizes its worst-case penalty (for any penalty function defined appropriately as before)
\emph{within this region}, ignoring the remaining unlikely cases.
Since \OurSys\ adopts general sampling methods for sensitivity analysis (\Cref{sec:sensitivity}) and finding robust plans (\Cref{sec:robust}),
they can be extended to compute highest-density regions and evaluate \Cref{eq:worst-in-hdr}.
Details and experimental evaluation are beyond the scope of this paper but would be interesting to pursue as future work.}

\section{Sensitivity Analysis: Morris Method}
\label{app:sensitive:morris}

Given a function $h: [0,1]^d \to \Reals$,
the \emph{Morris method}~\cite{morris1991factorial} uses a collection of $K$ seeds from the input domain
and calculates, for each seed, a derivative-based \emph{elementary effect} for each input dimension in a small neighborhood around the seed;
these elementary effects are then aggregated over all seeds to provide a sensitivity measure for each input dimension.
Specifically, starting with each seed $\sample{}$ (think of it as a point location in $[0,1]^d$),
the method picks a random ordering $\varsigma$ of the $d$ dimensions and generates a path
$\sample{} = \seq\sample0 \to \seq\sample1 \to \cdots \to \seq\sample{d}$ with $d$ steps, one for each dimension according to $\varsigma$.
Let $\varsigma(i)$ denote the ordinal position for dimension $i$ in $\varsigma$.
The step for dimension $i$, corresponding to $\seq\sample{\varsigma(i)-1} \to \seq\sample{\varsigma(i)}$,
would move the input point along dimension $i$ by some small step size $\Delta_i$,
while keeping all other coordinates unchanged;
in other words, $\seq\sample{\varsigma(i)-1}$ and $\seq\sample{\varsigma(i)}$ differ only in
$\seq{x_i}{\varsigma(i)-1} + \Delta_i = \seq{x_i}{\varsigma(i)}$.
The elementary effect of dimension $i$ on seed $\sample$ is calculated as
$\EE_i(\sample) = (h(\seq\sample{\varsigma(i)}) - h(\seq\sample{\varsigma(i)-1}))/\Delta_i$.
Finally, from the collection of $K$ seeds $\sample_1, \ldots \sample_K$,
we calculate the \emph{Morris-sensitivity} for dimension $i$ as $\smash{1 \over K} \sum_{j=1}^K |\EE_i(\sample_j)|$.

To apply this method in our setting to understand the plan $\plan$ obtained under selectivity estimates $\hat \sels$,
we analyze the function $h(\sels) = \Penalty(\plan, \sels)$ by drawing the $K$ seeds randomly by $f(\sels | \hat \sels)$.
This approach directs Morris to focus more on the more relevant regions of the selectivity space around $\hat \sels$.
Even though each individual elementary effect is derivative-based and local,
tallying them over all paths explored by Morris intuitively paints an overall picture of variability over the relevant regions.
There is a chance that Morris may still miss some critical features of $\Penalty(\plan, \sels)$;
we set the step size ($\Delta_i$) and $K$ large enough to mitigate this issue.
Our current implementation adopts the same setting of $\Delta_i = 0.05 \times \hat\sel_i$ for all dimensions,
but an interesting alternative worth investigating as future work would be to set step size for each dimension
according to the marginal error distribution for that dimension.

We denote the \emph{Morris-sensitivity} for dimension $i$ as $\sensimorris_i(\plan, \hat \sels)$.
Overall, this analysis uses $K$ seeds, each requiring evaluating $\Penalty$ $d+1$ times.
Each invocation of $\Penalty(\plan, \sels)$ requires calling $\Opt$ to obtain the optimal plan (and its cost) at $\sels$,
calling $\Cost$ to re-cost $\plan$ at $\sels$.
Hence, the total cost of Morris is $O(Kd)$ $\Opt$ and $\Cost$ calls.
We show practical $K$ values to reach convergence in \Cref{sec:expr}.

\begin{figure*}
    \centering
    \includegraphics[scale=0.105]{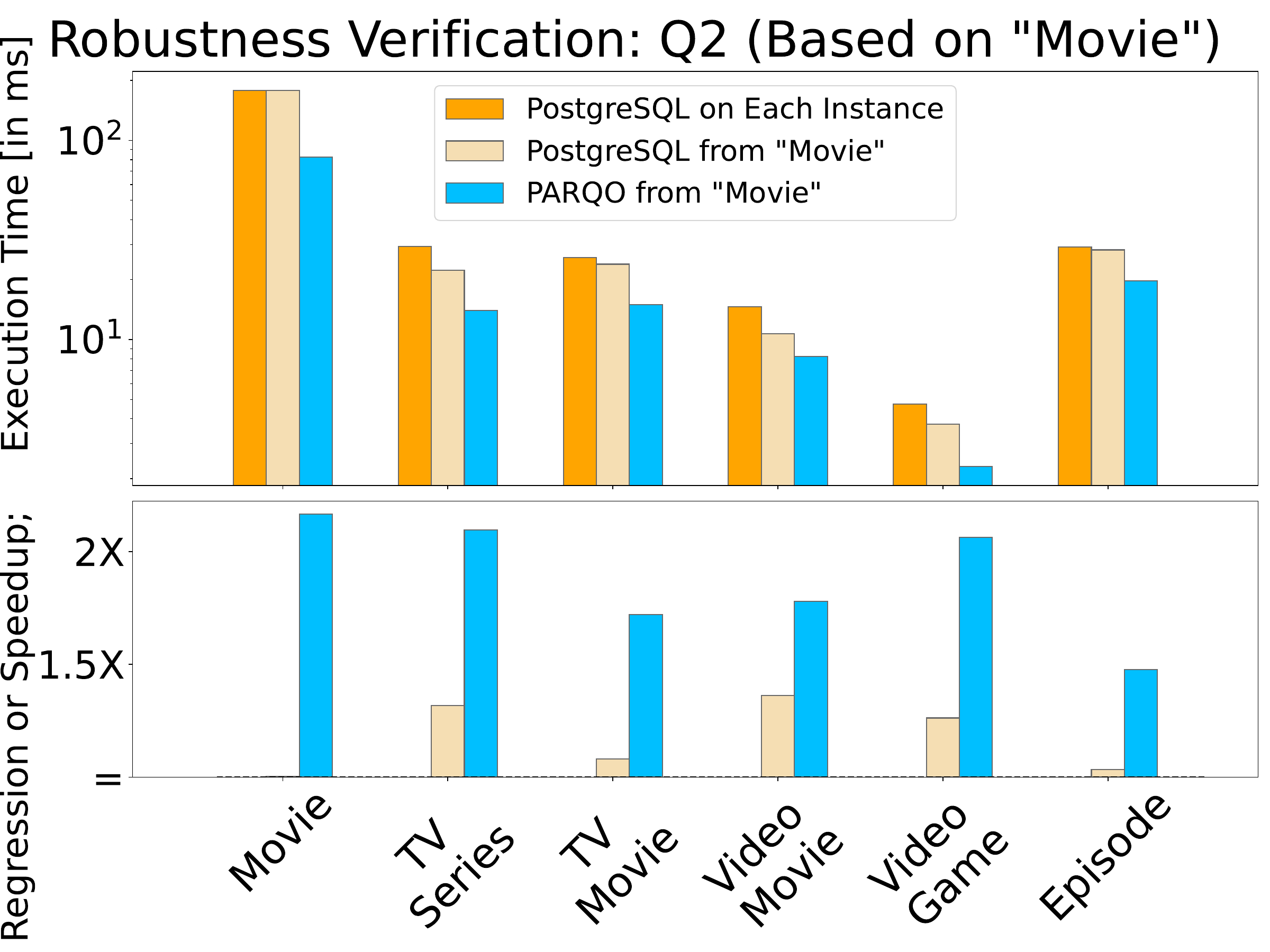}
    \ \ 
    \includegraphics[scale=0.105]{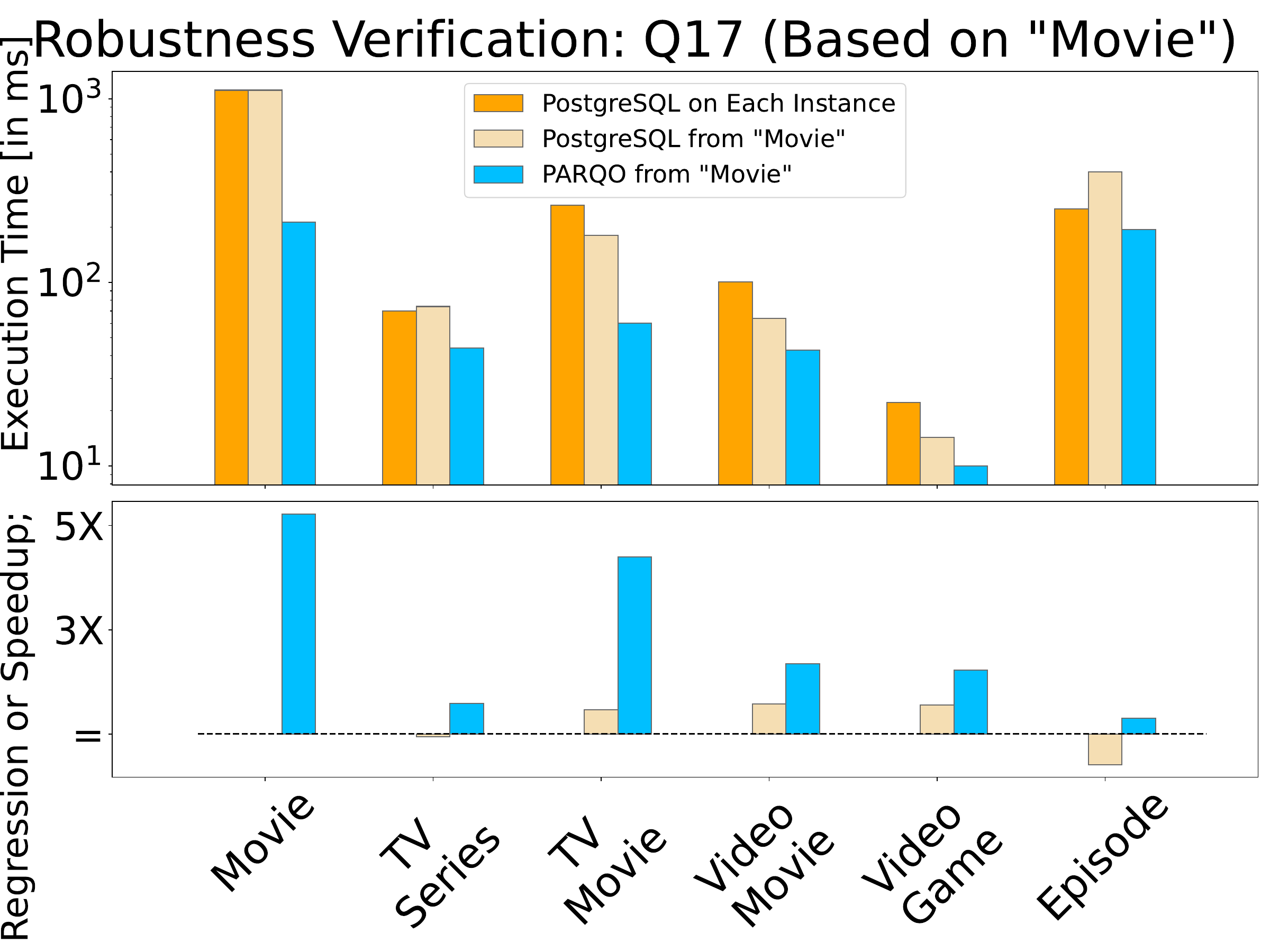}
    \ \
    \includegraphics[scale=0.105]{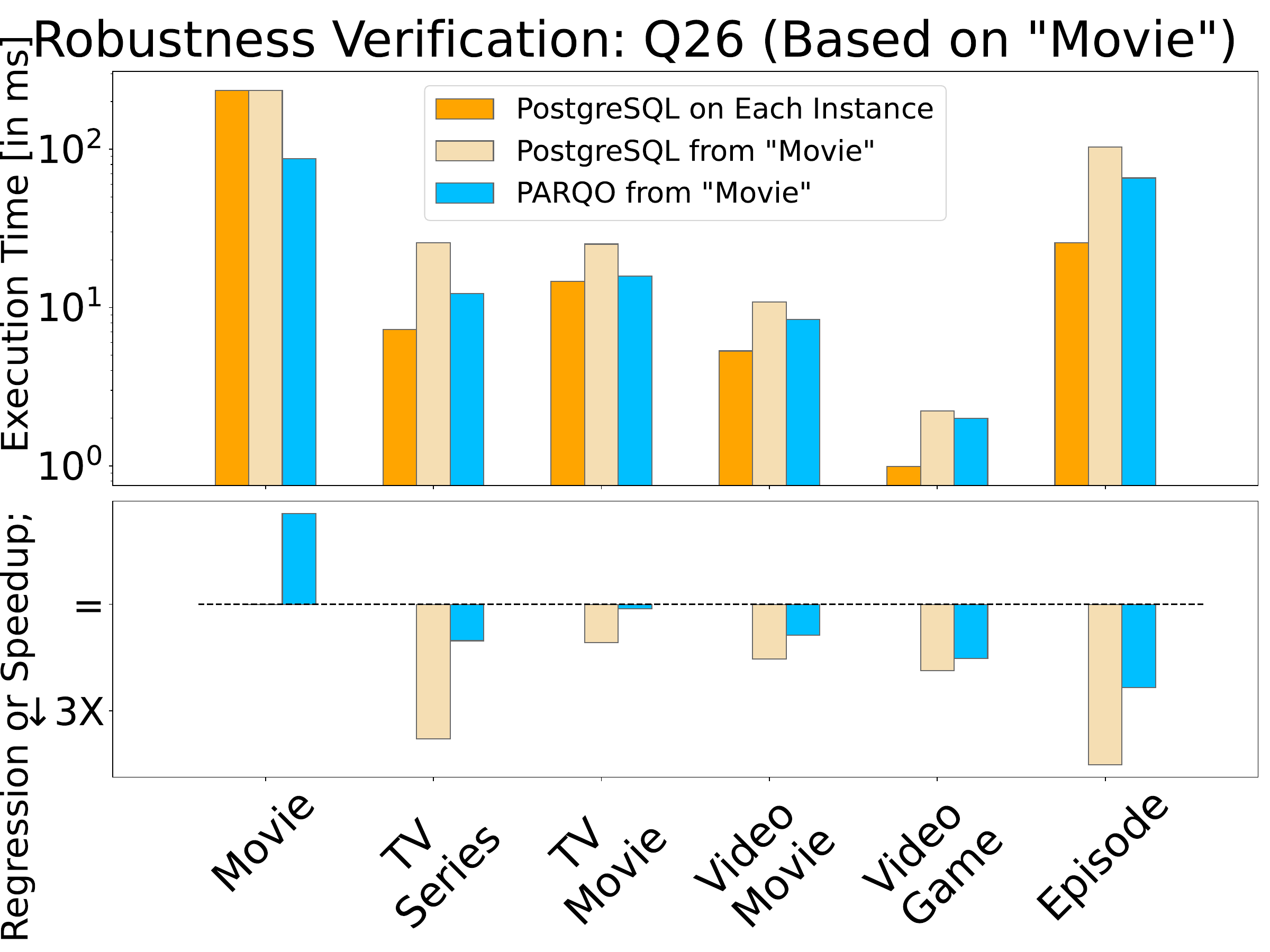}
    \ \
    \includegraphics[scale=0.105]{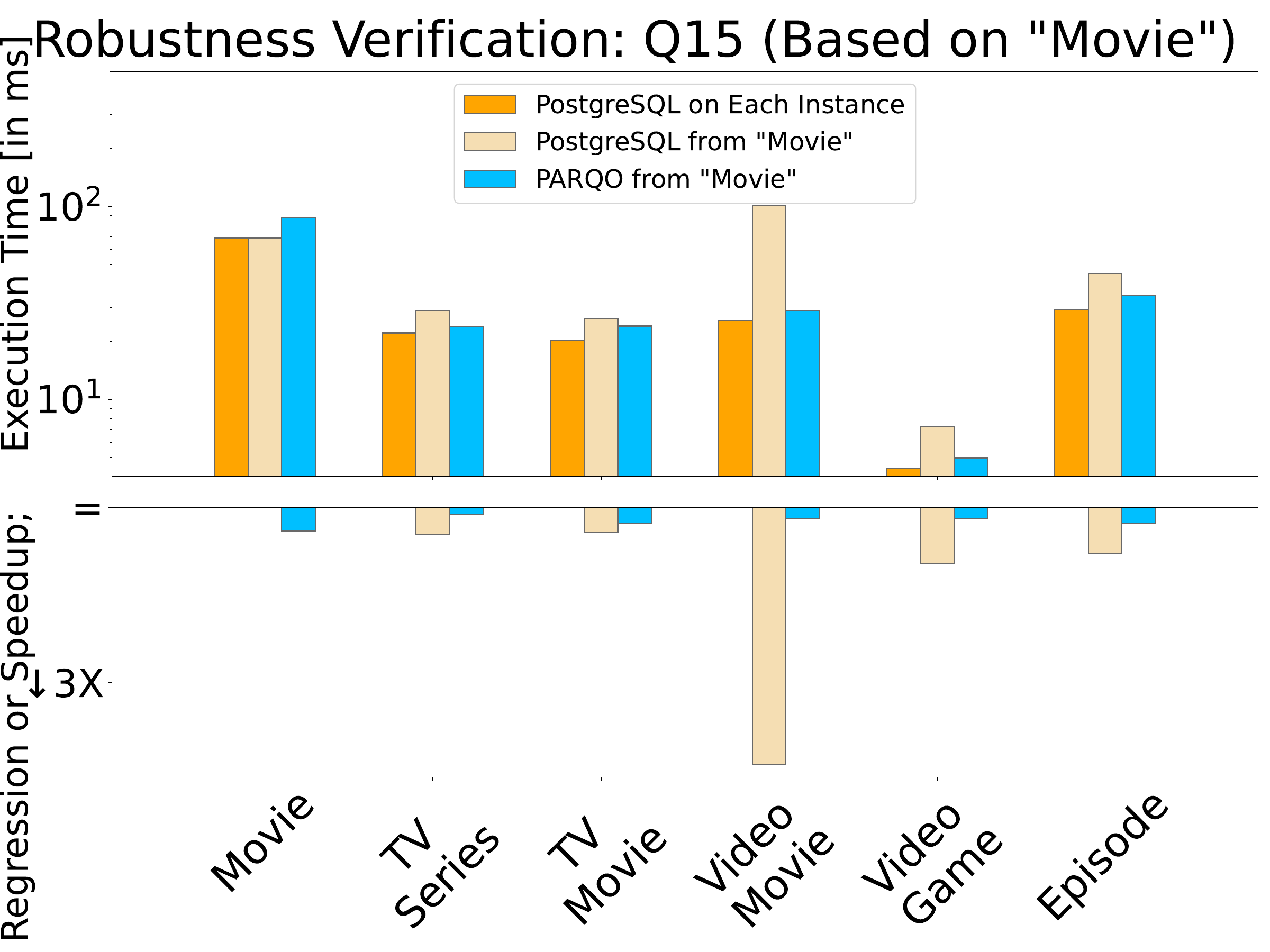}
    \vspace{-4mm}
    \caption{\small{\common{Actual execution times of JOB queries on multiple instances by category-slicing IMDB. ``Movie'' is the base instance.}}}
    \label{fig:verify-category-movie}
    \vspace{-3mm}
\end{figure*}

\begin{figure*}
    \centering
    \includegraphics[scale=0.105]{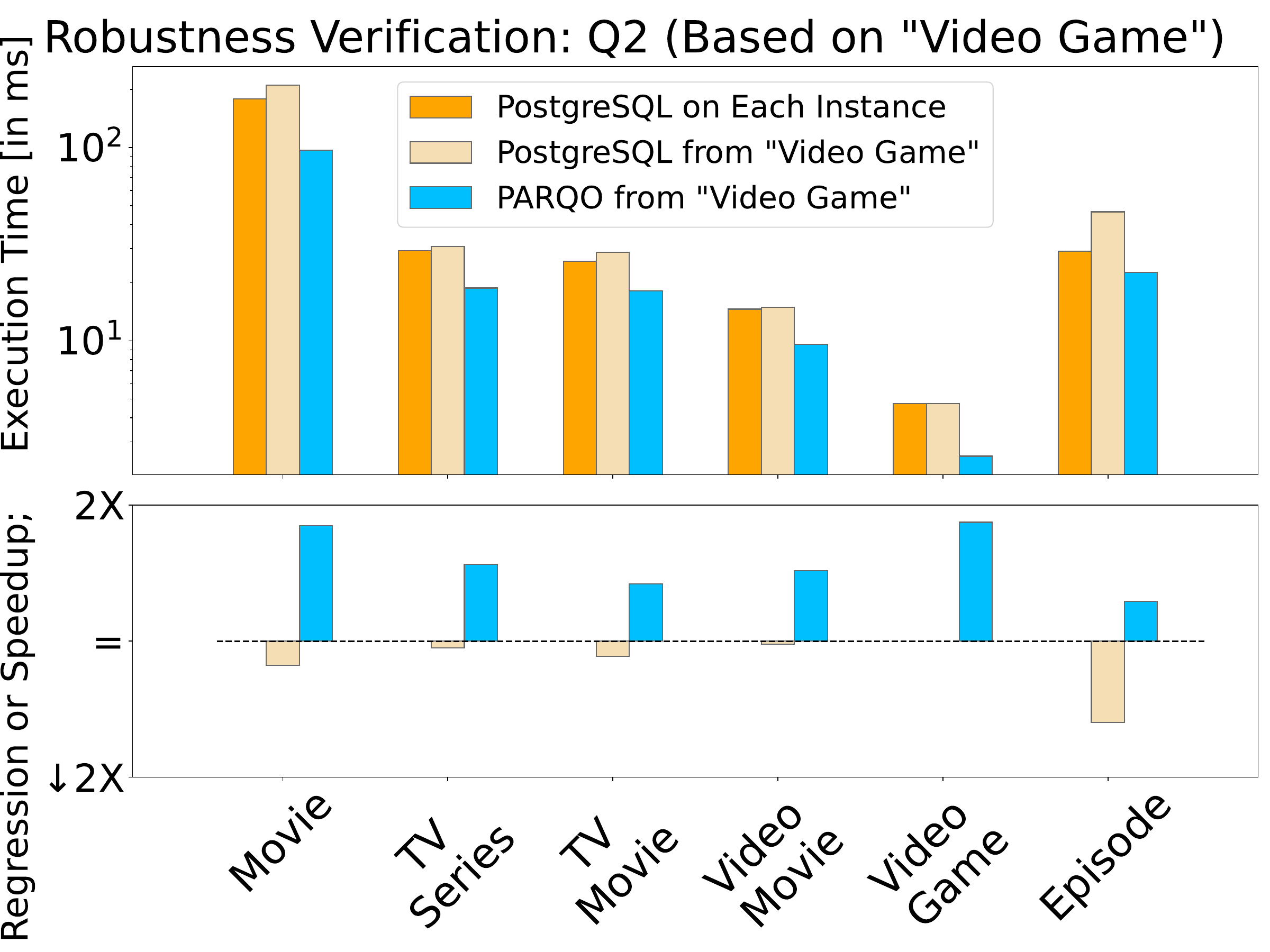}
    \ \ 
    \includegraphics[scale=0.105]{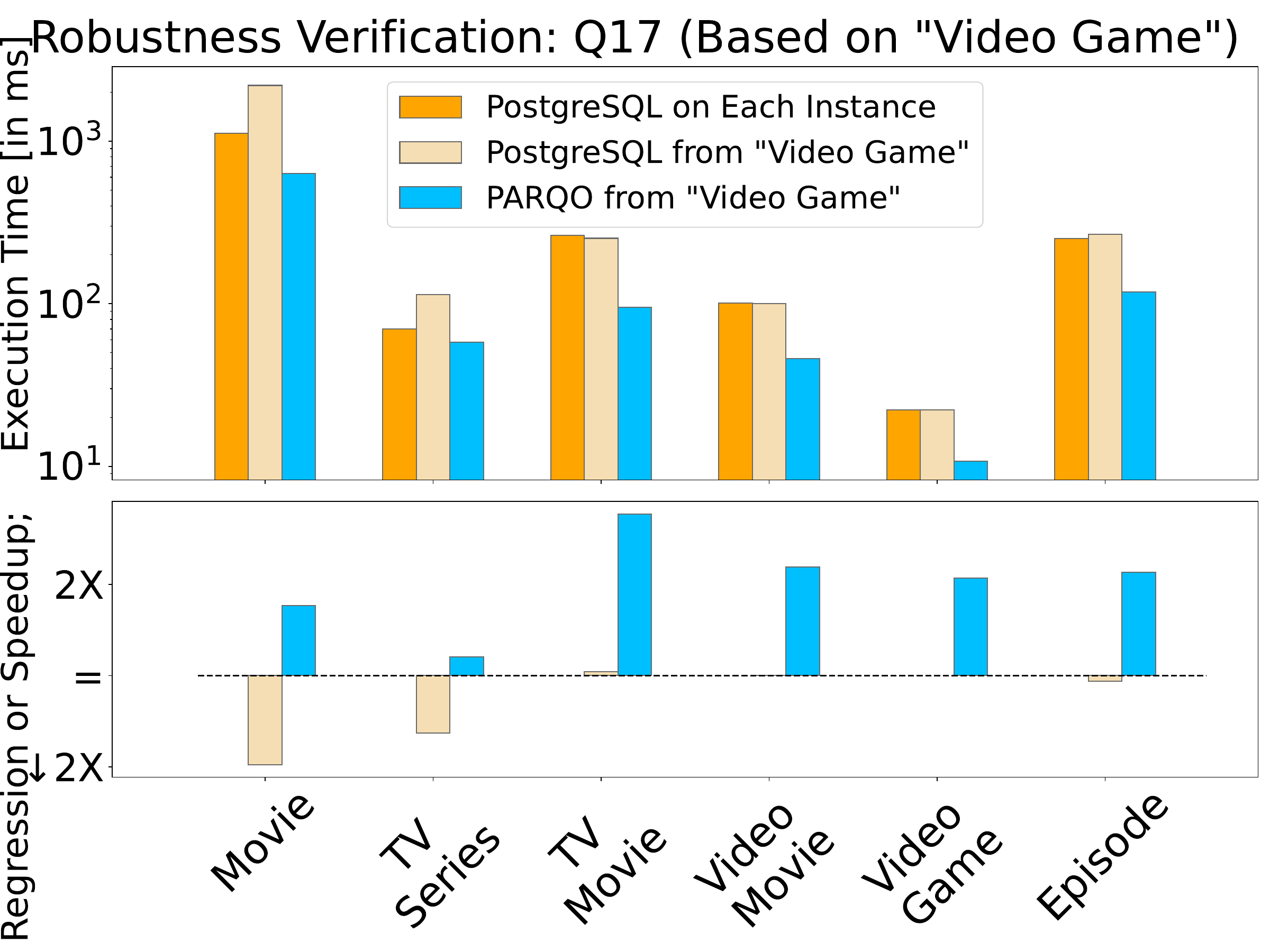}
    \ \
    \includegraphics[scale=0.105]{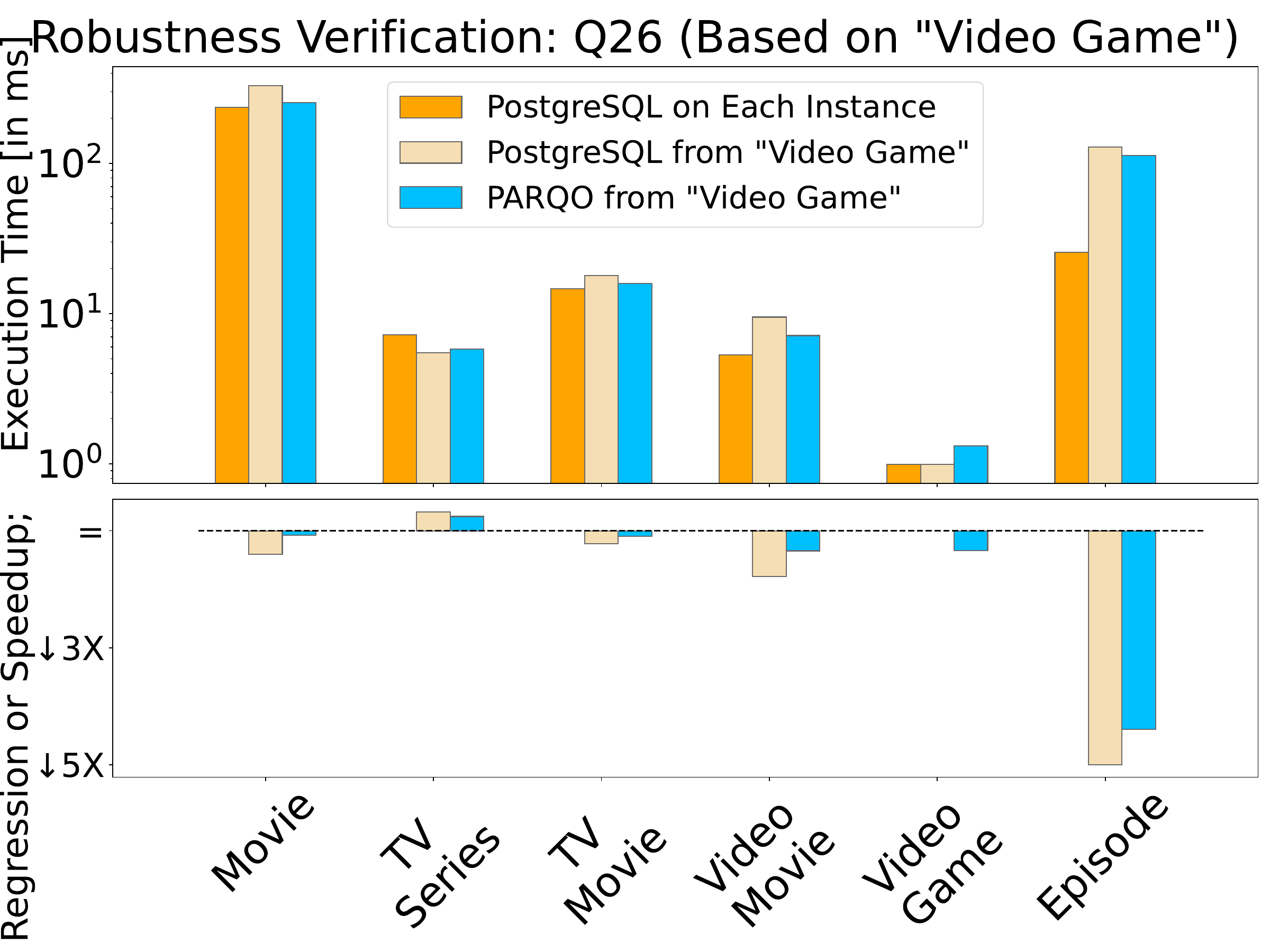}
    \ \
    \includegraphics[scale=0.105]{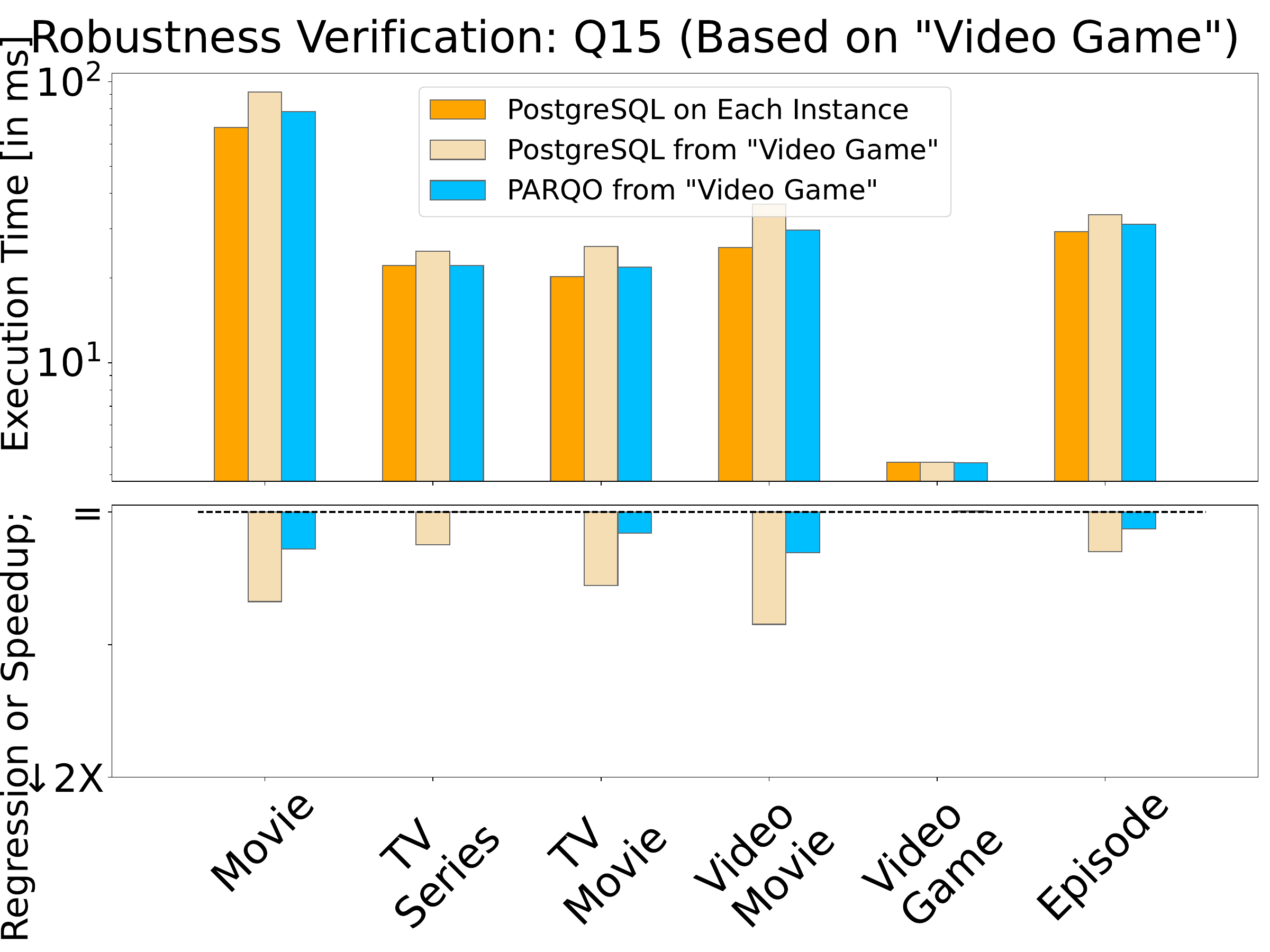}
    \vspace{-3mm}
    \caption{\small{\common{Actual execution times of JOB queries on multiple instances by category-slicing IMDB. ``Video Game'' is the base instance.}}}
    \label{fig:verify-category-video}
    \vspace{-2mm}
\end{figure*}

\section{Additional Results of Verifying Robustness }

\Cref{fig:verify-category-movie} summarizes the results when using ``Movie'' as the base instance%
\footnote{Q26 includes a predicate ``\textit{kt.kind = `movie'}'' on the \textit{kind\_type} table.
Under category-slicing, we modify it to ``\textit{kt.kind IS NOT NULL}'' to avoid returning empty results for some instances, which would not be useful.}
(additional results lead to similar conclusions and can be found in~\cite{fullversion}).
\OurSys\ does well for Q2 and 17, even outperforming the instance-optimized PostgreSQL plans across instances.
For Q26 and 15, \OurSys\ is slower than instance-optimized PostgreSQL plans in most cases, by not by much.
In comparison, the base PostgreSQL plan regresses much further in most cases;
Q15 on the ``Video Movie'' instance is particularly illustrative of \OurSys's robustness.

Besides using ``Movie'' as the base instance, which has the largest average number of rows among the instances partitioned by category, we additionally regard "Video Game" as the base instance, representing the smallest one with the smallest average number of rows. The other instances except ``Video Game'' serve as testing instances. 
From the results shown in Figure \ref{fig:verify-category-video}, we draw similar conclusions: for Q2 and Q17, \OurSys~ consistently outperforms both instance-optimized PostgreSQL plans and base PostgreSQL plans across instances. For Q15 and Q26, although not as effective as instance-optimized PostgreSQL plans, \OurSys~'s plan successfully avoids significant regressions.


\begin{figure}
    \begin{subfigure}[left]{0.48\linewidth}
        \includegraphics[scale=0.13]{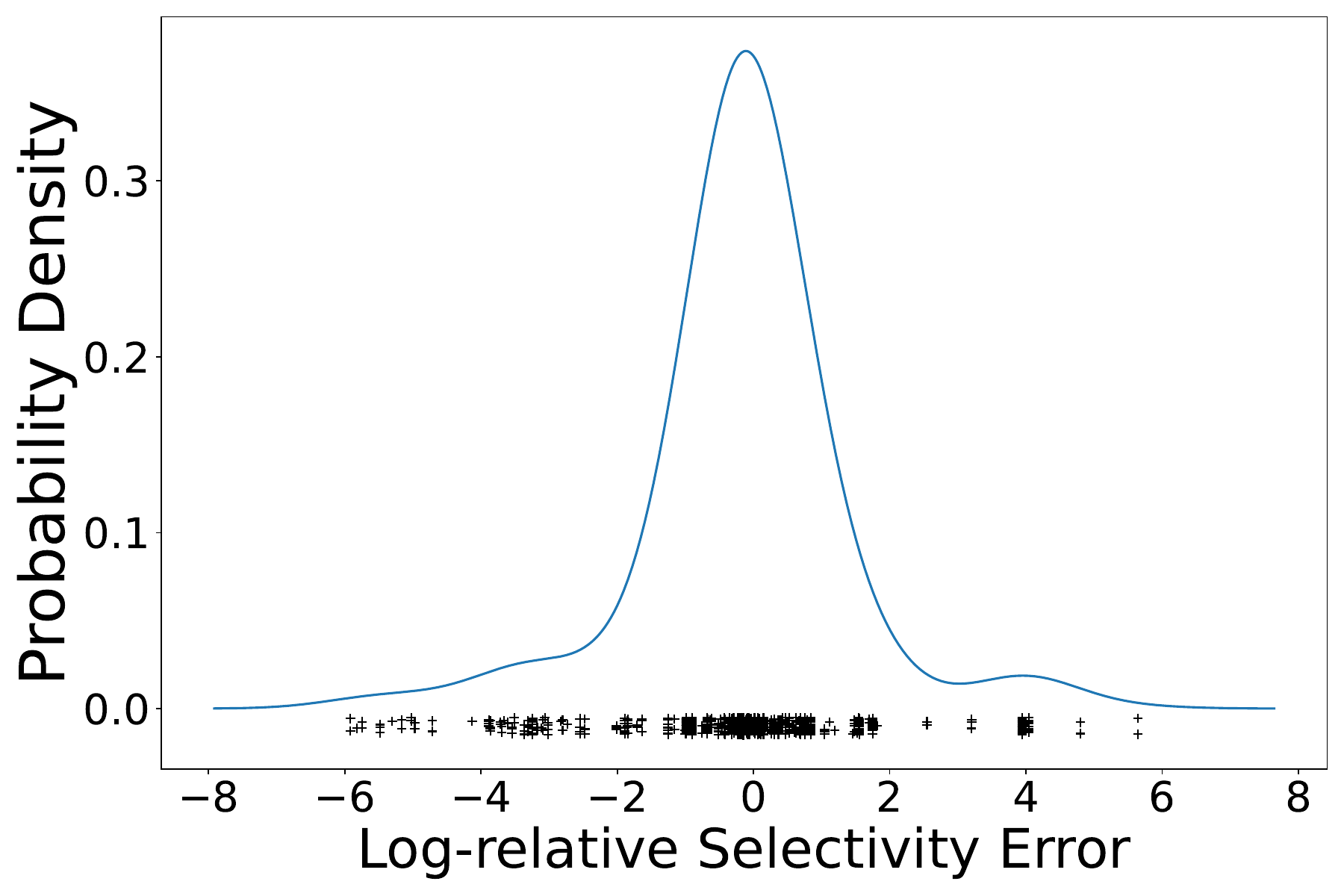}
        \vspace{-2mm}
        \caption{$mc^\select \join cn^\select$}
        \label{fig: mc-cn}
    \end{subfigure}
    \vspace{-3mm}
    \begin{subfigure}[right]{0.48\linewidth}
        \includegraphics[scale=0.13]{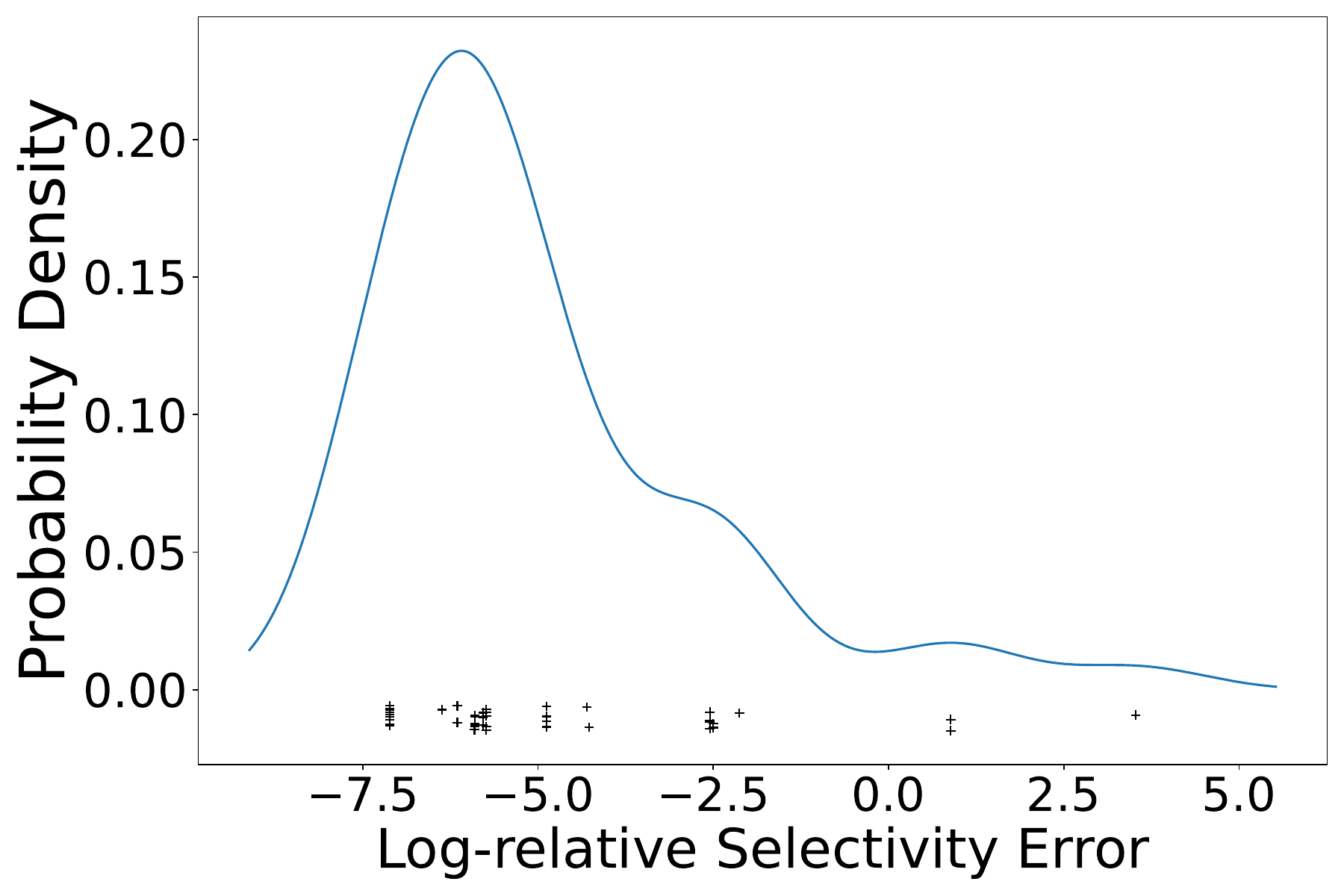}
        \vspace{-2mm}
        \caption{$mk \join k^\select$}
        \label{fig: mk-k}
    \end{subfigure}
    \vspace{-1mm}
    \caption{\small{\common{Examples error distributions for querylets in JOB, construct based on the profiling workload, which includes 80 instances from JOB. The dots in the plot represent the collected log-relative selectivity errors.}}}
    \label{fig:err distribution}
\end{figure}

\begin{figure*}
    \centering
    \includegraphics[scale=0.14]{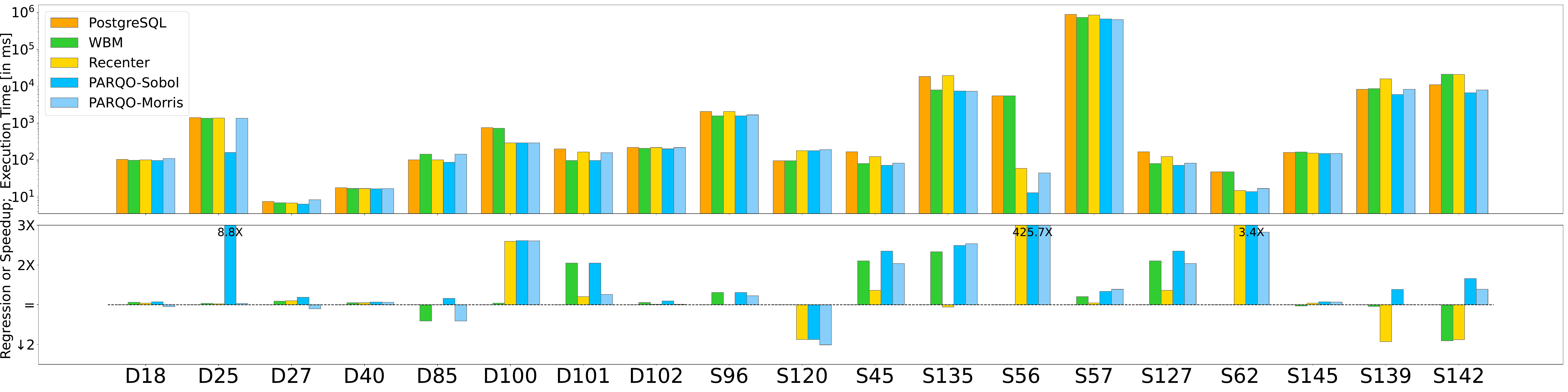}
    \caption{\small{
    \common{Actual execution times for different plans selected by various methods on DSB and STATS. Queries for which they perform similarly (7 out of 15 for DSB, and 15 out of 26 for STATS) are omitted.}}}

    \label{fig:rqo-dsb}
\end{figure*}

\begin{figure*}
    \centering
    \includegraphics[scale=0.14]{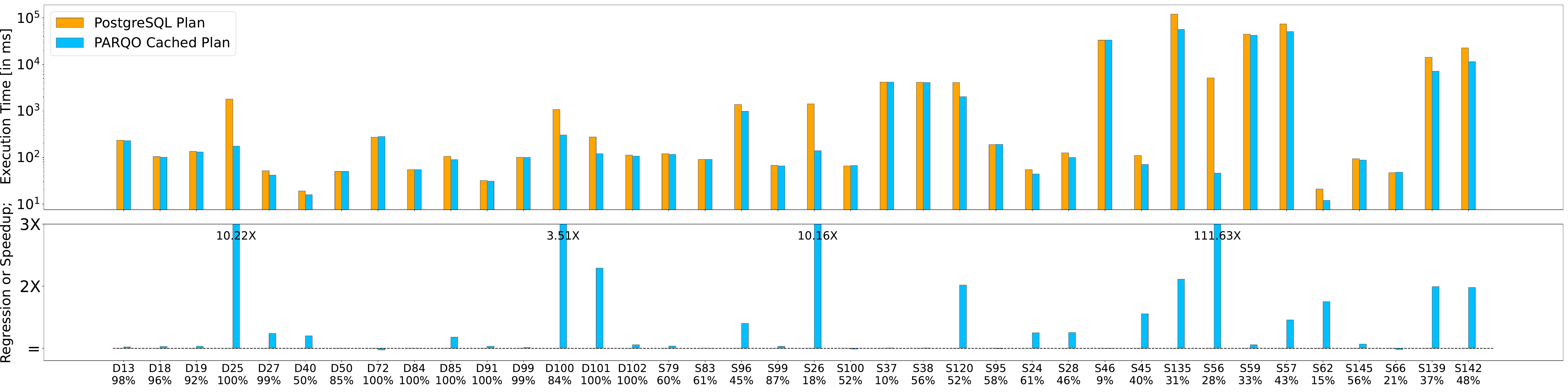}
    \caption{\small{
\common{PQO results for DSB and STATS benchmarks. The horizontal axis displays the query template ID and the reuse fraction of the template. 
}}}
    \label{fig:pqo-dsb}
\end{figure*}

\common{\section{Experiments on DSB and STATS}
\label{app:expr}

\mparagraph{DSB}
DSB~\cite{ding2021dsb} is a recently published industrial benchmark, positioned as an enhanced version of TPC-DS~\cite{nambiar2006making}. It incorporates more complex distributions and correlations, both for single columns and multi-dimensional data, within a table and across tables. As in \cite{wu2024modeling}, we focus on the 15 SPJ query templates, use a scale factor of 2 to populate the data, and apply the default settings to generate the query workload. For each query template, we generate 100 queries as the profiling workload and use a different random seed to create the workload for evaluation.
\mparagraph{STATS}
The data for STATS(-CEB)~\cite{han2021cardinality} is sourced from the Stack Exchange dataset. Its query workload includes 146 handpicked queries. Since there are no explicitly defined templates, we first randomly select 26 queries as the evaluation set, with the constraint that these query templates range from simple 3-table joins to more complex 7-table joins to ensure generalizability.
We extract 22 unique query templates from the evaluation workload.
We then use the remaining 120 queries as the profiling workload, constructing error profiles using the same method as for JOB, as discussed in \Cref{sec:expr}.

First, to compare the robust plan selected by \OurSys~ with other approaches for a single query, 
we use the same hyper-parameters introduced in \Cref{sec:expr}. 
Figure \ref{fig:rqo-dsb} shows the results for DSB and STATS. 
For queries in DSB, \OurSys-Sobol outperforms PostgreSQL in 8 out of 15 queries with an overall speedup of $3.43\times$. 
For the remaining 7 queries (omitted here), all methods select plans with nearly same runtime performance compared with PostgreSQL. 
Similar to our observation in JOB, \OurSys-Sobol outperforms both WBM and \OurSys-Morris in most cases. The overall speedup of \OurSys-Sobol across all DSB queries is $2.01\times$. 
Notably, for complex queries such as D100, 101, and 102, which contain non-PKFK many-to-many and non-equi joins, \OurSys-Sobol performs well, demonstrating its ability to discover a more robust plan for complex selectivity uncertainties by successfully capturing and penalizing them. 

We note similar behavior in the STATS benchmark. 
\OurSys-Sobol accelerates 10 out of 26 queries with an average of $1.36\times$ speedup than PostgreSQL. Since the execution time per query ranges from 10ms to 20 minutes, here we highlight the speedup achieved for each individual query as shown on the bottom of Figure \ref{fig:rqo-dsb}. Notably, for S56, PostgreSQL and WBM plans take more than 5,500 ms, whereas \OurSys-Sobol's plan only takes 13 ms, resulting in a $425.7\times$ speedup over PostgreSQL and WBM. \OurSys-Sobol shows regression on only one query, S120, with a $\downarrow1.87\times$, however we can still demonstrate the benefits later in the PQO experiments. PostgreSQL takes nearly 15.9 minutes to execute the entire testing query set, while our method completes it in 11.7 minutes, achieving an overall $1.36\times$ speedup. Again, since we are calculating the average speedup as mentioned in \cref{sec:expr}, this number is contributed more by those slower queries.

We use the same experimental setup for PQO as described in \Cref{sec:expr} and present the execution times in Figure \ref{fig:pqo-dsb}. The other performance measurements can be found in \ref{tab:performance_metrics}. In DSB, across all query instances, \OurSys~ achieves a 0.68-hour improvement compared with PostgreSQL, which takes 1.23 hours running all queries, resulting in a $2.24\times$ speedup. 
We also observe that the average reuse fraction in DSB is much higher than in the other benchmarks. After optimized one single query, \OurSys~ can benefit 93\% of new queries on average. This is because DSB defines the workload distribution for each query template, leading to more \textit{similar} queries generated from the same distribution. 
In contrast, most predicates in JOB and STATS are distinct and divergent. Therefore, the selectivities of some anchors differ from the majority from the evaulation workload.
For STATS, the average reuse fraction is 43\%\footnote{\common{Since there can be at most two queries per template in STATS, in Figure \ref{fig:pqo-dsb}, we only present results using the query with the higher reuse fraction as the anchor. But this reported average reuse fraction is calculated by the rates of triggering reuse for all anchors.}}, and the execution time saved across all 22{,}000 query instances is 11.47 hours, which is an $1.14\times$ speedup compared with PostgreSQL's 91.6 hours.
\footnote{\common{This execution time does not include those long-running queries where both the robust plan and PostgreSQL's plan exceeded the timeout threshold, set at 20 minutes in our experiments. When PostgreSQL's plan ran over 20 minutes while the robust plan finished within 20 minutes, we simply recorded PostgreSQL's latency as 20 minutes. We did not observe any cases where PostgreSQL's plan finished within 20 minutes but \OurSys's plan did not.}} 
For template S120, despite previous regression in RQO settings, \OurSys~ achieves a $2.02\times$ speedup in 52\% of generated new instances. Additionally, for templates S26, 24, 28, and 135, although the robust plans did not improve the single query performance as in Figure\ref{fig:rqo-dsb}, \OurSys~ achieves significant performance gains when applied to the query templates.
The optimization time also shows a speedup for these two benchmark. \OurSys~ achieves 0.09-hour and 0.01-hour improvements in the query optimization time for DSB and STATS respectively. Combining the improvements made by \OurSys in execution and optimization with the up-front overhead of \OurSys, which is 0.99 hours for DSB and 0.74 hours for STATS, the benefit becomes evident.
}

\end{document}